\documentclass[numberedappendix]{emulateapj}

\usepackage{graphicx}
\usepackage{epsfig}
\usepackage{amssymb}
\usepackage{amsmath}
\usepackage{graphicx}
\usepackage{ifpdf}
\usepackage{natbib}
\usepackage{bm}
\usepackage{xcolor}
\usepackage{comment}
\usepackage{mnsymbol}
\usepackage[export]{adjustbox}

\definecolor{darkgreen}{rgb}{0.0,0.5,0.0}
\usepackage[colorlinks,citecolor=darkgreen]{hyperref}

\newcommand{\FigDir}{Figures}
\newcommand{\const}{{\mbox{const.}}}
\newcommand{\ie}{\emph{i.e.,} }
\newcommand{\eg}{\emph{e.g.,} }
\newcommand{\cf}{\emph{cf.} }
\newcommand{\kappamax}{\kappa_{\rm max}}
\newcommand{\Gammamax}{\Gamma_{\rm max}}

\newcommand{\FullTitle}{{Nonlinear electromagnetic-wave interactions in pair plasma: (I) Non-relativistic regime}}
\newcommand{\ShortTitle}{{Nonlinear waves in $\mathrm{e}^\pm$ plasma}}

\begin{document}

\title{\FullTitle}
\shorttitle{\ShortTitle}
\shortauthors{Ghosh et al.}
\author{Arka Ghosh$^{\dagger}$}
\author{Daniel Kagan}
\author{Uri Keshet}
\author{Yuri Lyubarsky}
\affil{Physics Department, Ben-Gurion University of the Negev, POB 653, Be'er-Sheva 84105, Israel}
\thanks{$^{\dagger}$Electronic address: arka@post.bgu.ac.il}	
\date{\today}

\begin{abstract}
This paper is the first in a series devoted to the numerical study of nonlinear interactions of electromagnetic waves with plasma. We start with non-magnetized pair plasmas, where the primary processes are induced (Compton) scattering and the filamentation instability. In this paper, we consider waves in which electron oscillations are non-relativistic. Here, the numerical results can be compared to analytical theory,  facilitating the development of appropriate numerical tools and framework. We distill the analytic theory, reconciling plasma and radiative transfer pictures of induced scattering and developing in detail the kinetic theory of modulation/filamentation instability. We carry out homogeneous numerical simulations  using the particle-in-cell codes EPOCH and Tristan-MP, for both monochromatic waves and wave packets. We show that simulations of both processes are consistent with theoretical predictions, setting the stage for analyzing the highly nonlinear regime.
\end{abstract}

\date{Accepted ---. Received ---; in original ---}
\label{firstpage}


\maketitle

\section{Introduction}
The interaction of intense radiation with plasma is of great interest both for laboratory physics and for astrophysics. In the laboratory, the main interest is in studying the propagation of short, narrow laser pulses through a dense electron-ion plasma, in which case the main effects are self-focusing,  induced Raman scattering, and the generation of wake-fields due to charge separation (see, e.g., the review by \citealt{Mourou_etal06}).
In astrophysical systems, the spatial and time scales of the radiation are macroscopically large; even the millisecond pulses may be considered as macroscopic as compared with the wave period. In this case, self-focusing yields to filamentation of the radiation flux: the wide beam splits to narrow subbeams. In most astrophysical applications, the radiation frequency is well above the plasma frequency, in which case the non-linear effects become weaker. Nevertheless, even weak non-linear corrections could accumulate considerably over a long enough path, thus significantly affecting the propagation of the wave. In the high frequency, low density regime, induced Raman scattering becomes weaker than induced Compton scattering. Moreover, in many of the systems of interest, the radiation propagates through an electron-positron (henceforth pair) plasma, rendering charge separation effects, including Raman scattering, irrelevant.
Therefore, for astrophysical applications, induced Compton scattering and the filamentation instability are the most important processes.

In the modern plasma literature, induced Compton scattering is sometimes called stimulated Brilluin scattering (e.g., \citealt{Edwards_etal16, Schluck_etal17}). The term Brilluin scattering implies a three-wave interaction of electromagnetic and sound waves. The latter do not exist in non-magnetized, collisionless plasmas (with the exception of electron-ion plasmas with electrons much hotter than ions), in which the electromagnetic waves are scattered off density fluctuations. This process is called the induced Compton scattering both in classical plasma works (e.g., \citealt{Litvak_Trachtengerts71, Galeev_sunyaev73,Drake_etal74,Ott_etal74,Lin_Dawson75}) and in the astrophysical literature. Therefore, we use this term (henceforth referred to simply as induced scattering), retaining the term Brilluin scattering for scattering off ion sound waves in two-temperature plasmas or off magnetosonic waves in magnetized plasmas.

In the astrophysical literature, the effect of induced scattering was considered in active galactic nuclei \citep[AGN;][]{Syunyaev71,Coppi_etal93},  pulsars \citep{Blandford_Scharlemann76,Wilson_Rees78,Gedalin_Eichler93,Thompson94,LyubarskY_Petrova96,Petrova08b,Petrova08a,Petrova09}, interstellar masers \citep{Galeev_sunyaev73,Montes77}, and fast radio bursts \citep[FRB;][]{Lyubarsky08,Lyubarsky20,Lyubarsky_Ostrovska16,Lu_Kumar18,Metzger19,Beloborodov20,Kumar_Lu20}. In the special case where the electron oscillations in the field of the wave are relativistic, induced scattering was studied by \citet{Lyubarsky19a} .
These works focused on the conditions for the escape of waves, by analytically estimating the effective optical depth.

The filamentation instability attracted much less interest in the astrophysical community, presumably because this effect is unable to prevent the escape of the waves.
Nevertheless, the instability may affect the properties of the outgoing radiation \citep{Sobacchi_etal20}. Namely, the filamentation instability implies the formation of narrow subbeams, which are strongly diffracted when the radiation escapes the plasma slab. This could lead to smearing of short bursts in time and/or to frequency modulations of the observed intensity due to interference between the subbeams.  Filamentation was observed in numerical simulations of relativistic, magnetized shocks, in which case a strong electromagnetic precursor is emitted upstream \citep{Iwamoto_etal17,Sironi_etal21}.

Our goal is to study the observational features of the above nonlinear processes, by simulating numerically the propagation of intense radiation through a pair plasma.  In the case of induced scattering, this opens a possibility not only to estimate the optical depth, but also to find the characteristics of the outgoing radiation. In the case of the filamentation instability, we are interested in the quantitative parameters of the sub-beams, which can facilitate quantitative predictions for the time-frequency structure of the observed radiation. Among other sources, these effects are of special interest for FRBs. Their extremely high brightness temperature implies that close enough to the source, the electrons oscillate relativistically in the field of the FRB radiation \citep{Luan_Goldreich14}. A quantitative study of the above effects in such an extreme case is possible only numerically.

In this paper, we develop the setup for particle-in-cell (PIC) simulations of the propagation of intense radiation through pair plasmas, and demonstrate both induced scattering and the filamentation instability. Both  processes may be considered as an instability of the pumping beam. Here, we consider only the case where the electron oscillations in the field of the wave are not relativistic. In this regime, the linear stage of the instability may be considered analytically, so we can verify the simulation results. The analysis of the highly nonlinear evolution, including the case of truly strong waves, in which electron oscillations become relativistic, is deferred for a future publication.
We present the theoretical background in \S\ref{sec:theory}, the PIC simulation setup in \S\ref{sec:simulations}, and the results in \S\ref{sec:results}. We summarize and discuss our findings in \S\ref{sec:summary}.
We use Gaussian cgs units with $k_B=c=1$.

\section{Nonlinear  electromagnetic-wave interactions with pair plasmas}

\label{sec:theory}

The propagation of electromagnetic waves is conveniently studied using the wave equation for the vector potential $\bm{A}$,
\begin{equation}
    \frac{\partial^2 \bm{A}}{\partial t^2}-\Delta\bm{A}=4\pi\bm{j}\,.
\end{equation}
In this paper, we consider a nonrelativistic pair plasma. Here, the current is presented as
\begin{equation}
    {\bm j}=\sum qN\bm{v}\, ,
\label{eq:current}\end{equation}
where the summation is over the species, $q=\pm \mathrm{e}$, $\mathrm{e}$ is the electron charge, $N$ is the density of the species, and $\bm{v}$ is the particle velocity.

The particle oscillations in the field of the wave are found from the equation of motion,
\begin{equation}
    m\frac d{dt}\frac{\bm{v}}{\sqrt{1-v^2}}=q\left[-\frac{\partial\bm{A}}{\partial t}+\bm{v\times (\nabla\times A)}\right]\,,
\label{eq:motion}\end{equation}
where $m$ is the electron mass.
In this paper, we assume that the oscillation velocity is nonrelativistic, $v\ll 1$. Then to first order in the small parameters  $v$ and $\mathrm{e}A/m$,
\begin{equation}
    {\bm v}^{(1)}=-\frac qm\bm{A}\,.
\label{eq:velocity} \end{equation}
The unperturbed plasma is assumed to be neutral, with a particle number density $N_{\rm tot}=2N_0$, where $N_0$ is the unperturbed electron density.
In the first approximation, Eq.~(\ref{eq:current}) then becomes
\begin{equation}
   4\pi \bm{j}^{(1)}=-\omega_p^2\bm{A}\, ,
\label{linear_currrent}\end{equation}
where
\begin{equation}
    \omega^2_p=\frac{4\pi \mathrm{e}^2N_{\rm tot}}m = \frac{8\pi \mathrm{e}^2N_0}m
\end{equation}
is the plasma frequency squared. 

Now the wave equation may be written as
 \begin{equation}
\frac{\partial^2\bm{A}}{\partial t^2}-\Delta \bm{A}+\omega_p^2\bm{A}=4\pi \bm{j}^{\rm
nl}\, ,
 \label{wave_eqn1}\end{equation}
where $\bm{j}^{\rm nl}$ is the nonlinear current, which could arise  from non-linear corrections to the electron velocity or/and from variations of the plasma density caused by the wave. In our approximation, the nonlinear current is small, so the nonlinear corrections are small at the wavelength scale. However, when the nonlinear current is in resonance with the wave, the corrections accumulate, and so could lead to a significant modification of the initial wave at longer time scales.

\subsection{Induced scattering}
\label{subsec:InducedScattering}

Even though the term "induced scattering" refers to the quantum picture, the effect is purely classical: the beating between the initial and the scattered waves produces a ponderomotive force, which gives rise to a plasma density modulation, which in turn leads to an enhanced scattering rate. With such an approach, the process can be considered as an instability: when the plasma is illuminated by a pumping radiation beam, the amplitudes of both the scattered wave and the density modulation grow.  The derivation of the induced
scattering of electromagnetic waves in plasma was given by, e.g., \citet{Litvak_Trachtengerts71} and \citet{Drake_etal74}. Here, we outline the derivation in the simplest case of a pair plasma, and find the condition under which the plasma theory results reduce to those of radiation transfer theory.

\begin{figure}
    \centering
    \includegraphics[width=0.49\textwidth,trim={4.5cm 9cm 7cm 3.5cm}, clip]{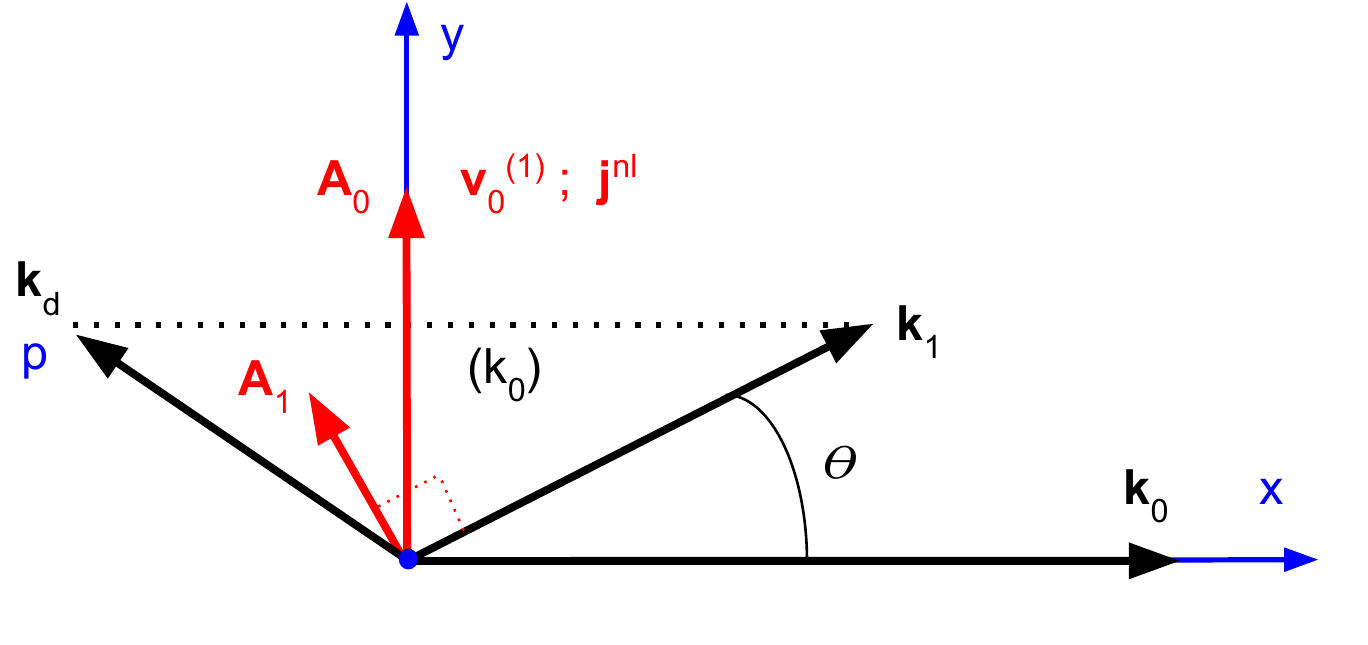}
    \caption{Illustration of our induced-scattering setup.}
    \label{fig:illust}
\end{figure}

Having in mind 2D simulations, we consider scattering in the $xy$ plane, with the initial wave propagating along the $x$-axis and the scattered wave being offset by an angle $\theta$, assuming both waves being polarized in the $xy$ plane (see illustration in Fig.~\ref{fig:illust}; the generalization to a general setup is discussed farther below). Then the electromagnetic vector potential may be written as
\begin{equation}
    {\bm A}={\bm A}_0e^{i(k_0x-\omega_0 t)}+{\bm A}_1e^{ik_1(x\cos\theta+y\sin\theta)-i\omega_1 t}+{\rm c.c.}\,,
\label{eq:A-potential}\end{equation}
where indices 0 and 1 refer respectively to the initial and scattered waves, $\bm{A}_0$ and $\bm{A}_1$ are in the $xy$ plane, c.c. is the complex conjugate, and $\bm{r}=\{x,y,z\}$ are Cartesian coordinates.

The beating wave, of angular frequency and wave vector given respectively by
\begin{equation}
    \omega_d=\omega_1-\omega_0 \quad \mbox{and} \quad \bm{k}_d=\bm{k}_1-\bm{k}_0\, ,
\label{eq:beating}\end{equation}
is in Landau resonance with particles moving with the phase velocity of the wave, $\bm{v}=\omega_d\bm{k}_d/k^2_d$. These particles "see" a force constant in time that produces a density modulation,
\begin{equation}
    \delta N \,e^{i\bm{k}_d\cdot\bm{r}-i\omega_dt}+{\rm c.c.} \,,
\label{eq:fluct}\end{equation}
with an amplitude growing in time. Therefore, the rate of scattering on these fluctuations grows, too. Thus the mechanism redistributes the energy of the initial wave into the energies of the scattered wave and the plasma particles. An important point is that inasmuch as $v\ll 1$ and the scattering angle is typically large, $k_d\sim k_0$ and $\omega_d\sim \omega_0 v\ll\omega_0$.

The density perturbations may be derived from the Vlasov equation,
\begin{equation} \label{eq:Vlassov}
    \frac{\partial f}{\partial t}+\bm{v\cdot\nabla}f+\bm{F}\cdot\frac{\partial f}{\partial \bm{p}}=0\,,
\end{equation}
where $f$ is the distribution function, $\bm{p}$ is the particle momentum, and $\bm{F}$ is the force exerted on the particles by the beating wave.
Inasmuch as the beating frequency $\omega_d$ is small as compared with $\omega_0$ and $\omega_1$, one could express such a slowly varying force via the ponderomotive potential (e.g., \citealt{Schmidt79}),
\begin{equation}
 \bm{F}=-\bm{\nabla}\varphi;\qquad   \varphi\equiv\frac{\mathrm{e}^2\langle{\bm A}^2\rangle}{2m}\, ,
\label{ponderomotive}\end{equation}
where the angular brackets denote averaging over a time period longer than $\omega_0^{-1}$ but short as compared with $\omega_d^{-1}$. Substituting Eq.~(\ref{eq:A-potential}) in Eq.~(\ref{ponderomotive}) yields $\varphi=\mbox{const.}+\varphi_d\exp(i\mathbf{k}_d\cdot \mathbf{r}-i\omega_d t)+\rm c.c.$, where
\begin{equation}
   \varphi_{d}
    =\frac{\mathrm{e}^2}{m}{\bm A_1}\cdot\bm A_0^*=\frac{\mathrm{e}^2}{m}\mu  A_1A_0^* \, ,
\label{eq:pond}\end{equation}
and $\mu\equiv \cos\theta$.

The ponderomotive potential is a quadratic function of the particle charge, therefore the electron and positron densities remain equal even in the presence of fluctuations.
Looking for perturbations of the form (\ref{eq:fluct}), we may consider the 1D Vlasov equation for the distribution of particle velocities in the direction $\bm{\hat{k}}_d\equiv \bm{k}_d/k_d$. One can conveniently normalize the 1D distribution function $f(p)$ by unity, 
\begin{equation}
    \int_{-\infty}^{\infty} f\,dp=1\,,
\end{equation}
where $p\equiv \bm{p}\cdot\bm{\hat{k}}_d$, and integrals are henceforth carried out over their full parameter range. In the oscillating ponderomotive potential, the initial particle distribution function, $f_0$, gets an oscillating perturbation of the form $\delta f\exp(i\mathbf{k}_d\cdot\mathbf{r}-i\omega_d t)+\rm c.c.$ Linearising Eq.~(\ref{eq:Vlassov}) in the small parameters $\delta f/f$ and $\varphi/\epsilon$, where $\epsilon\equiv p^2/2m$ is the particle energy, one gets
\begin{equation}
    (\omega_d-vk_d)\delta f=-k_d \varphi_{d}v\frac{df_0}{d\epsilon}\,.
\label{eq:Vlasov_lin}\end{equation}
The density perturbations are now found as
\begin{equation}
\!\!\frac{\delta N}{N_0}=\int \delta f\,dp=\frac{k_d\mathrm{e}^2 \mu}{m}A_1A_0^*\int\frac{v\, d f_0/d\epsilon}{k_dv-\omega_d-i0}\,dp\,,
\label{eq:DeltaN}\end{equation}
where the Landau rule for bypassing the pole was used, with $i0$ designating an infinitesimal positive imaginary constant.
Now we can find the evolution of the scattered wave.

The wave equation (\ref{wave_eqn1}) for the scattered wave becomes
\begin{equation}
    (k_1^2-\omega_1^2+\omega_p^2)A_1=4\pi \bm{\hat{A}}_1 \cdot \bm{j}^{\rm nl}_1,
\label{scattered_wave}\end{equation}
where $\bm{j}_1^{\rm nl}$ is the component of the non-linear current  having the angular frequency $\omega_1$ and the wave vector $\bm{k}_1$, and $\bm{\hat{A}}_1\equiv \bm{A}_1/A_1$ is the unit vector in the direction of $\bm{A}_1$. 
The nonlinear current is produced by beating between the low frequency density fluctuations (\ref{eq:fluct}) and the velocity oscillations in the field of the pumping wave. According to Eq.\ (\ref{eq:velocity}), these velocities are just $\bm{v}_0^{(1)}=-(q/m)\mathbf{A}_0$, so 
\begin{equation}
    4\pi\bm{j}^{\rm nl}_1=4\pi \sum q\,\delta N\,{\bm v}_{0}=-\omega_p^2\frac{\delta N}{N_0}{\bm A}_0\,.
\label{eq:nl_curr}\end{equation}
Substituting this current into Eq.\ (\ref{scattered_wave}), one obtains the equation for the scattered wave,
\begin{equation}
    \left(k_1^2+\omega_p^2-\omega_1^2\right)A_1=
    -\omega_p^2\mu\frac{\delta N}{N_0}A_0\,.
\label{eq:nonlinear_wave}\end{equation}
Plugging in $\delta N$ from Eq.\ (\ref{eq:DeltaN}), the wave equation turns into the dispersion equation
\begin{equation}
    k_1^2+\omega_p^2-\omega_1^2=
    -\frac{(\mathrm{e}\omega_p\mu)^2}{m} \vert A_0\vert^2\int\frac{k_dv\,df_0/d\epsilon}{k_dv-\omega_d-i0}\,dp\, .
\label{eq:dspr}\end{equation}
Instead of the incident wave amplitude, one can conveniently use the dimensionless strength parameter
\begin{equation}
    a \equiv\frac{2\mathrm{e}\vert\bm{A}_0\vert}{m}\,,
\label{eq:a}\end{equation}
where the factor 2 appears because our definition (\ref{eq:A-potential}) implies that the wave amplitude is $2A_0$. According to Eq.\ (\ref{eq:velocity}), the condition that the oscillation velocity in the field of the wave is non-relativistic now becomes $a\ll 1$.
Using $a$, the dispersion equation becomes
\begin{equation}
    k_1^2+\omega_p^2-\omega_1^2=
    -\frac 14m(\omega_pa\mu)^2 \int\frac{k_dv\,df_0/d\epsilon}{k_dv-\omega_d-i0}\,dp\,. \label{eq:scatdisp}
\end{equation}
The scattering rate is determined by the imaginary part of $\omega_1$. The wave $\omega_0$ is scattered into $\omega_1$, so that $A_1$ grows, if ${{\rm Im\,}\omega_1>0}$. In the opposite case, $A_1$ decreases as the wave $\omega_1$ is scattered into $\omega_0$. The imaginary part of $\omega_1$ is determined by the imaginary part of the integral in the RHS of the dispersion equation, which arises because of the pole in the integrand. Therefore, only particles in Landau resonance with the beating wave are responsible for the effect. 

Generally, the dispersion equation is solved numerically, as we show later. However, if we consider the growth rate to be small enough,
\begin{equation}
\vert{\rm Im}\,\omega_1\vert\ll \left(\frac{T}m\right)^{1/2}k_d\,, \label{eq:conditionscat}
\end{equation}
where $T$ is the plasma temperature in energy units, we can  use the Sokhotski-Plemelj formula in the RHS of the dispersion equation.
Then, approximating $\vert{\rm Im}\,\omega_1\vert\ll\omega_1$ in the LHS,
we obtain
\begin{equation}
    {\rm Im}\,\omega_1=\frac{\pi}8\frac{(\omega_p a\mu m)^2}{\omega_1}\frac{\omega_1-\omega_0}{\vert\bm{k}_0-\bm{k}_1\vert}\left(\frac{df_0}{d\epsilon}\right)_{v=\frac{\omega_1-\omega_0}{\vert\bm{k}_0-\bm{ k}_1\vert}}\,.
    \label{eq:DispersionA}
\end{equation}
As $df_0/d\epsilon<0$, we see that if $\omega_1<\omega_0$, then ${\rm Im}\,\omega_1>0$, i.e., the pumping wave $\omega_0$ is scattered into the wave $\omega_1$.
Thus, the wave frequency decreases due to induced scattering.

In astrophysical applications, one typically defines the rate of induced scattering as the growth rate of the scattered wave energy,  $\kappa=2{\rm Im}\,\omega_1$.
Substituting the Maxwell distribution for $f_0$ and taking into account that
\begin{equation}
    \vert{\bm k}_0-{\bm k}_1\vert^2=(\omega_0-\omega_1)^2+
    2(1-\mu)\omega_0\omega_1\approx 2(1-\mu)\omega_0\omega_1\,,
    \label{eq:kdLLomegad}
\end{equation}
one finds
\begin{gather}
    \kappa=\frac{\pi^{1/2}(\omega_0-\omega_1)(\omega_p a \mu)^2}{8\omega_0^{1/2}\omega^{3/2}_1\sqrt{1-\mu}}\left(\frac mT\right)^{3/2}
    e^{-\frac m{4T}\frac{(\omega_0-\omega_1)^2}{\omega_0\omega_1(1-\mu)}}.
    \label{eq:finalgrowth}
\end{gather}
This result coincides with the induced  scattering rate used in radiation transfer theory \citep{Zeldovich_etal72}. The term $\mu^2$ in the numerator applies to scattering in the plane; it should be replaced by $(1+\mu^2)/2$ for the typical radiation-transfer case of 3D scattering of non-polarized radiation.

\begin{figure}[h!]
    \centering
    \includegraphics[width=0.49\textwidth]{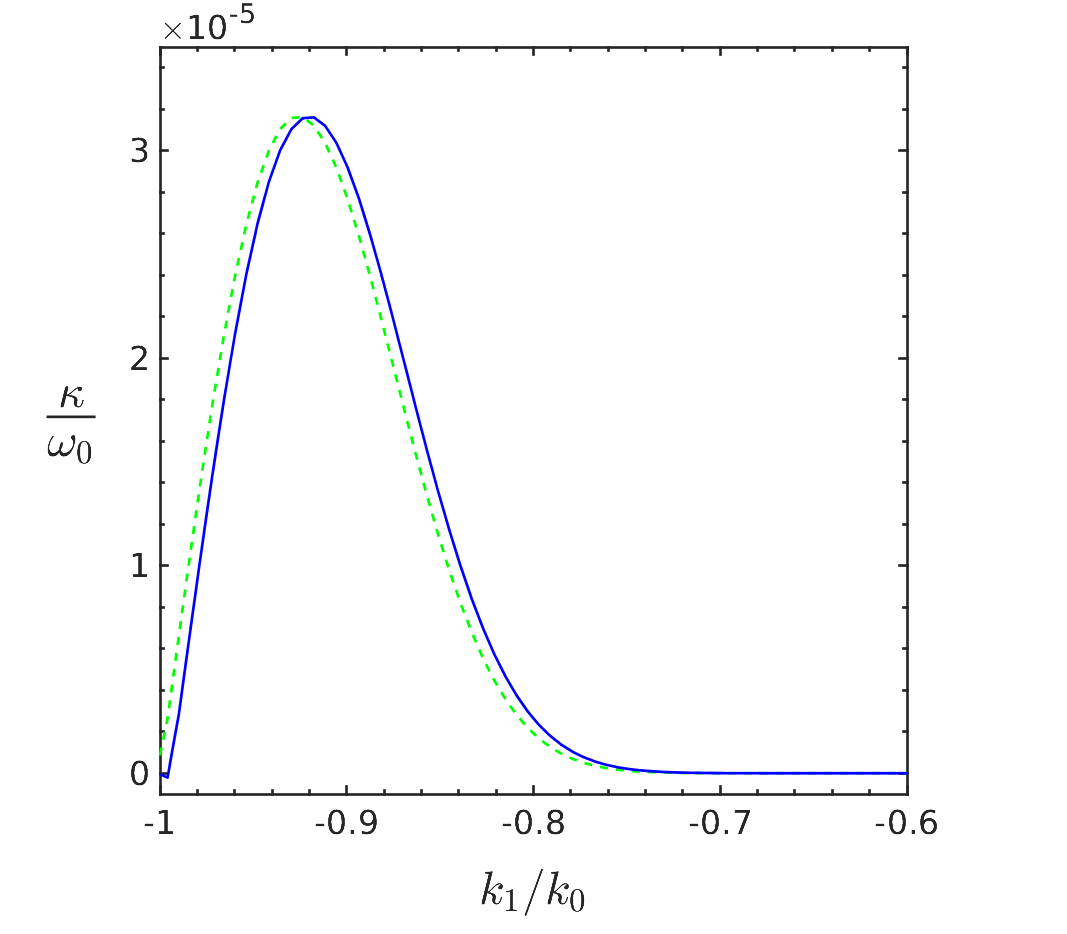}
    \caption{
    The induced-scattering (energy growth) 
    rate $\kappa = 2 \text{Im}(\omega_1)$,
    normalized to the incident angular frequency, evaluated by numerical integration of the dispersion relation \eqref{eq:scatdisp} (solid blue) and according to the asymptotic solution \eqref{eq:finalgrowth} (dashed green), which is valid under the condition \eqref{eq:scatter_validity}. Parameters:
    $T/m = 0.00137$, $\omega_p/\omega_0 = 0.107$, $a = 0.0043$, $\theta = \pi$,
    and a 1D Maxwellian $f_0$. 
    \label{fig:numscat}}
\end{figure}

Equation (\ref{eq:finalgrowth}) indicates that the growth rate is maximal for backward scattering, $\mu=-1$; in fact, this applies for the general case of 3D scattering with arbitrary polarization. Induced scattering outside the emission beam occurs due to a weak background radiation (produced, e.g., by spontaneous scattering). In this case, the energy of the scattered radiation grows exponentially, $\propto \exp{(\kappa t)}$. Many e-folding times are necessary in order to scatter a significant fraction of the beam energy. Therefore, the beam is scattered predominantly into the states corresponding to the maximal scattering rate. The fastest growing mode has the angular frequency
\begin{equation}
  \omega_{1,\rm max}\simeq \left(1-2\sqrt{T/m}\right)\omega_0 \label{eq:isfreqshift}\,,
\end{equation}
the corresponding wave vector being
\begin{equation}
k_{1,\rm max}\simeq-\left(1-2\sqrt{T/m}\right)k_0\, .
\label{eq:Backscattering}
\end{equation}
Here, we denote $k_1=\hat{\bm{k}}_0\cdot \bm{k}_1$, so the minus sign indicates backscattering.
Substituting Eq.~\eqref{eq:isfreqshift} into Eq.~\eqref{eq:finalgrowth}, expanding around $T/m \to 0$ and retaining the leading order term yields the maximal scattering rate
\begin{equation}
\kappamax \simeq \sqrt{\frac{\pi}{32e}}\frac{\omega_p^2}{\omega_0}\frac{ma^2}T\, .
\label{kappa_scat_max}\end{equation}
The validity condition for the asymptotic estimate (\ref{eq:finalgrowth}), given by Eq.\ (\ref{eq:conditionscat}), now becomes
\begin{equation}
    a\ll 5
    \frac{\omega_0}{\omega_p}\left(\frac Tm\right)^{3/4}\,.
\label{eq:scatter_validity}\end{equation}
Figure \ref{fig:numscat} presents the growth rate of the backscattered wave, derived both from the numerical solution to the dispersion equation (\ref{eq:scatdisp}) and from the analytic asymptotics (\ref{eq:finalgrowth}). We find a good agreement between the two as long as Eq.~\eqref{eq:scatter_validity} is satisfied, which is indeed the case in the figure.
More examples are provided below, comparing PIC simulations of the scattering process to the theoretical estimates.

The above results were derived for a monochromatic pumping wave. For astrophysical applications, the case of a broad spectrum is more interesting. Assuming that the initial beam is composed of waves with different frequencies and random phases, the electromagnetic potential becomes (\cf\ Eq.\ (\ref{eq:A-potential}))
\begin{equation}
    \bm{A}=\sum_{\omega}\bm{A}_{0,\omega}e^{i(kx-\omega t)}+\sum_{\omega'}\bm{A}_{1,\omega'}e^{ik'(x\cos\theta+y\sin\theta)-i\omega' t}+\rm c.c.
\end{equation}
Generally, the summations should be carried out over $\bm k$ modes, but as the directions of both initial and scattered beams are fixed, suffice to sum over absolute values $k$, equivalent to summations over frequencies.
Now, Eq.\ (\ref{eq:pond}) for the ponderomotive potential is replaced by
\begin{equation}
    \varphi_d=\frac{\mathrm{e}^2}{m}\mu  \sum_{\omega}A_{1,\omega+\omega_d}A_{0,\omega}^*\,,
\end{equation}
which yields the density fluctuations (\cf\ Eq. (\ref{eq:DeltaN}))
\begin{equation}
\frac{\delta N_{\omega_d}}{N_0}=\frac{k_d\mathrm{e}^2 \mu}{m}\sum_{\omega}A_{1,\omega+\omega_d}A_{0,\omega}^*\int\frac{v\, d f_0/d\epsilon}{vk_d-\omega_d-i0}\,dp\,.
\label{eq:delta_N1}\end{equation}
The nonlinear current having angular frequency $\omega_1$ and wave vector $\bm{k}_1$ (\cf\ Eq.\ (\ref{eq:nl_curr})) now becomes
\begin{equation}
    4\pi\bm{j}^{\rm nl}_1=
    -\omega_p^2\sum_{{\tilde\omega}}\frac{\delta N_{\omega_1-{\tilde\omega}}}{N_0}{\bm A}_{0,{\tilde\omega}}\,.
\end{equation}
Averaging this current over random phases using Eq.\ (\ref{eq:delta_N1}), and plugging the result into Eq.\ (\ref{scattered_wave}), one gets a dispersion equation similar to Eq.\ (\ref{eq:dspr}), but with $\vert A_0\vert^2$ replaced by $\sum_{\omega}\vert A_{\omega}\vert^2$. The resulting induced-scattering rate in the slow-growth limit of Eq.~(\ref{eq:conditionscat}) is
\begin{equation}
    \kappa=\frac{\pi^{1/2}\omega_p^2m^{3/2}\mu^2}{8T^{3/2}\sqrt{1-\mu}}\int \frac{\vert a_{\omega}\vert^2(\omega-\omega_1)}{\omega^{1/2}\omega_1^{3/2}}
    e^{-\frac m{4T}\frac{(\omega-\omega_1)^2}{\omega\omega_1(1-\mu)}}d\omega,
\label{eq:kappa_spectr}\end{equation}
where the dimensionless amplitude, $a_{\omega}$, is related to $A_{\omega}$ just as $a$ is related to $A_0$; see Eq.\ (\ref{eq:a}). It will be shown below that the induced scattering rate of a wide spectrum beam is weaker than that of a monochromatic beams. Therefore the condition (\ref{eq:conditionscat}) is less restrictive in the wide spectrum case.

If the width of the spectrum is small compared to the width of the particle distribution, $\Delta\omega/\omega\ll v_T\simeq (T/m)^{1/2}$, 
scattering proceeds as for a monochromatic beam. In the opposite limit of a broad spectrum, the particle distribution function may be approximated by the delta-function, so the frequency shift in the scattering vanishes, $\omega_1\to\omega$. Then one can substitute in Eq.\ (\ref{eq:kappa_spectr}) the expression
\begin{equation}
  \frac{\omega-\omega_1}{4\pi^{1/2}}\left[\frac {m/T}{(1-\mu)\omega\omega_1}\right]^{3/2}e^{-\frac m{4T}\frac{(\omega-\omega_1)^2}{\omega\omega_1(1-\mu)}}\to -\frac{\partial\delta(\omega-\omega_1)}{\partial\omega}
  \end{equation}
and integrate by parts, which yields
\begin{equation}
    \kappa=\frac{\pi}{2}\omega_p^2\mu^2(1-\mu)\frac{\partial}{\partial\omega}\left(\omega\vert a_{\omega}\vert^2\right) \propto \frac{\omega |a_\omega|^2}{\Delta \omega}\,,
\label{eq:ind_scat_wide_spectr}\end{equation}
where $\omega$ is both the incident and scattered frequency. If the total energy in the pumping beam, $\sim \omega^2\vert a_{\omega}\vert^2\Delta\omega$, remains fixed, one sees that the scattering rate decreases with increasing spectral width, as $\kappa \propto \Delta\omega^{-2}$.

Astrophysical applications usually use the radiation transfer equation, for the radiation intensity $I(\nu,\mathbf{\hat n})$ in frequency $\nu$ and direction $\mathbf{\hat n}$. For the induced scattering of non-polarized emission with a wide spectrum, this equation is written as (e.g., \citealt{Wilson82})
\begin{gather}
   \frac{d I(\nu,\mathbf{\hat n})}{d t}=\kappa I(\nu,\mathbf{\hat n})\,;\\
\kappa=\frac{\mathrm{e}^4N_{\rm tot} }{m^3}
\int\frac{1+(\bm{\hat n}\cdot\bm{\hat n}')^2}2
(1-\bm{\hat n}\cdot\bm{\hat n}')\frac{\partial
}{\partial\nu}\frac{I(\nu,\mathbf{\hat n}')}{\nu}d\Omega'\,.
  \label{kinComp}
\end{gather}
Here, the time derivative is taken along the ray,  and the integral is over solid angle $\Omega$.
In order to compare Eqs.\ (\ref{eq:ind_scat_wide_spectr}) and \eqref{kinComp}, note that in our case, the incident radiation is directed along the $x$-axis, such that $\mathbf{\hat n}'\cdot\mathbf{\hat n}=\mu$ and the radiation flux is
\begin{equation}
    F=\int I(\nu,\mathbf{\hat n}')d\Omega' d\nu\,.
    \end{equation}
The radiation flux can also be expressed using the wave amplitudes as
\begin{equation}
    F=2\frac{\langle \mathbf{E}^2\rangle}{4\pi}=\frac{m^2}{4\pi \mathrm{e}^2}\int\omega^2\vert a_{\omega}^2\vert d\omega,
\end{equation}
where the factor 2 in the first equality takes into account both polarizations. Comparing the last two equations shows that in our case, $\int I(\nu,\mathbf{n}')d\Omega'$ should be replaced by $m^2\omega^2\vert a^2_{\omega}\vert /2\mathrm{e}^2$.
Then, Eq.\ (\ref{kinComp})  reduces to our Eq.\ (\ref{eq:ind_scat_wide_spectr}) if one also substitutes $(1+\mu^2)/2$ by $\mu^2$, in order to take into account that we considered the scattering of polarized radiation in the polarization plane.

\subsection{Filamentation and modulation instabilities}

Due to the nonlinear interactions of radiation and matter, the refraction index depends on the
radiation power, which could lead, under certain conditions, to self-modulation or/and self-focusing
of the radiation beam (see, e.g., \citealt{Karpman75}).
If the radiation beam is wide, which is typically the case in astrophysical applications, self-focusing leads to filamentation of the beam into a set of subbeams. Note that formally, the filamentation instability is a special case of the modulation instability, when modulation is perpendicular to the direction of the beam. However, the physics and the conditions for filamentation substantially differ from those for modulation in the longitudinal direction. Specifically in a pair plasma, only the filamentation instability is possible, as shown below.

The modulation of the amplitude of the pumping wave  could be described by introducing two sidebands satisfying
\begin{equation}
    \omega_{\pm}=\omega_0\pm {\rm Re}\,\omega_d+{\rm Im}\,\omega_d;\quad
    {\bm k}_{\pm}={\bm k}_0\pm{\bm k}_d\,.
\label{eq:conserv}\end{equation}
Here, $\omega_0$ and ${\bm k}_0$ are the angular frequency  and wave vector  of the pumping wave, respectively, and $\omega_d$ and ${\bm k}_d$  are the angular frequency and wave vector of modulation.
In order to study evolution of the modulation, one can consider the non-linear interaction of these waves \citep{Drake_etal74, Max_etal74}, as for induced scattering. Instead, we exploit the fact that in this case, both $\omega_d\ll\omega$ and $k_d\ll k_0$, so we deal with a slow, long-wave modulation of a plane, monochromatic wave; the generalization for a broad spectrum is discussed at the end of this section.  

The propagation of the incident wave in plasma is described by the wave equation (\ref{wave_eqn1}). The non-linear current  for a monochromatic wave is found in the high-frequency limit as (\citealt{Montgomery_Tidman64,Sluijter_Montgomery65}; see also Appendix \ref{sec:current}) 
 \begin{equation}
    4\pi\bm{j}^{\rm nl} =\frac 14\omega_p^2a^2\bm{A}, \label{eq:nl_current}\end{equation}
where $a^2\equiv 2\langle (eA/m)^2\rangle$ is the dimensionless amplitude of the wave; this definition is equivalent to Eq.\ (\ref{eq:a}). Taking into account that the non-linear current is directed along $\bm{A}$, Eq.\ (\ref{wave_eqn1}) reduces to an equation for a scalar $A$. For a slowly varying wave, one can still use the monochromatic non-linear current (\ref{eq:nl_current}), in which one has to substitute the local amplitude of the wave. Therefore, the wave equation takes the form
\begin{equation}
\frac{\partial^2A}{\partial t^2}-\Delta A+\omega_p^2A=\frac 14\omega_p^2a^2A\,,
\label{eq:nl_wave_eq}\end{equation}
where $a$ is now the slowly varying dimensionless amplitude of the wave. Note that in the field of a modulated wave, the plasma density is expected to be modulated; therefore, one has to use the local density in the plasma frequency.

Equation (\ref{eq:nl_wave_eq}) has an exact solution describing a planar, monochromatic wave in a homogeneous medium,
\begin{equation}
A=(m/e)a_0\cos(\omega_0t-k_0x)\,,
\label{eq:plane_wave}\end{equation}
where $a_0$ is a dimensionless constant,
with a nonlinear dispersion law
\begin{equation}
    \omega_0^2=k_0^2+\omega_p^2\left(1-\frac{a_0^2}4\right).
\label{eq:nl_dispersion}\end{equation}
Let us consider the stability of this wave with respect to a slow, long-wavelength modulation.

One can conveniently use the complex dimensionless amplitude, $a(\bm{r},t)$, deviating from $a_0$, 
such that
\begin{equation}
   \frac em A={\rm Re}\,\left[a(\bm{r},t)e^{-i(\omega_0t-k_0x)}\right].
\label{eq:complex_A}\end{equation}
Then the slowly varying, complex function $a(\bm{r},t)$ contains information on both the amplitude and the phase of the wave.
One must take into account that in Eqs.\ (\ref{eq:nl_current}) and (\ref{eq:nl_wave_eq}), $a$ is the dimensionless real amplitude of the wave, so in the new notations, one has to replace $a^2$ by $\vert a\vert^2$. An important point is that the nonlinear evolution of the wave occurs not only due to the non-linear current (\ref{eq:nl_current}), but also because the plasma is expelled by  the ponderomotive force from the regions of enhanced radiation intensity, which affects the refraction index \citep{Drake_etal74}. Therefore, we substitute $\omega^2_p$ by $\omega_p^2\left(1+\delta N/N_0\right)$ in the wave equation \eqref{eq:nl_wave_eq} (but not in the dispersion equation \eqref{eq:nl_dispersion}, which is written for an unperturbed density), so henceforth $\omega_p$ denotes the plasma frequency for the unperturbed density.
Now the wave equation takes the form
\begin{align}
&2i\omega_0\left(\frac{\partial a}{\partial t}+v_g\frac{\partial a}{\partial
x}\right)-\frac{\partial^2 a}{\partial
t^2}+\Delta a \nonumber\\
&-\omega_p^2\left(1-\frac{\vert a\vert^2}4\right)\frac{\delta N}{N_0}a+\frac{\omega_p^2}{4}
(\vert a\vert^2-a_0^2)a=0,
\label{eq:WaveEqFora}
\end{align}
where
\begin{equation}
\bm{v}_g=\frac{d\omega_0}{d\bm{k}_0}=\frac{\bm{k}_0}{\omega_0}
\label{eq:vGroup}
\end{equation}
is the group velocity of the plane monochromatic wave, and we assumed, without loss of  generality, that the constant $a_0$ is real and positive.

In order to derive the evolution of a small modulation of the plane wave, we express the wave amplitude as
$ a=a_0+\delta a(\bm{r},t)$,
where $\vert\delta a\vert\ll a_0$.
Linearising the wave equation  with respect to $\delta a/a_0$ and $\delta N/N_0$ and neglecting $\vert a\vert^2$ with respect to unity in the fourth term,
we get
\begin{gather}
2i\omega_0\left(\frac{\partial \delta a}{\partial t}+v_g\frac{\partial \delta a}{\partial
x}\right)-\frac{\partial^2 \delta a}{\partial
t^2}+\Delta\delta a \nonumber\\
-\omega_p^2a_0\frac{\delta
N}{N_0}+\frac{\omega_p^2}{4} a_0^2(\delta a+\delta a^*)=0.
\label{wave_eqn}\end{gather}
The  modulation of the incident wave (\ref{eq:plane_wave}) may be found by substituting into this equation the ansatz
\begin{equation}
    \delta a=a_+e^{i(-\omega_dt+\bm{k}_d\cdot\bm{r})}+
    a_-^*e^{-i(-\omega_d^*t+\bm{k}_d\cdot\bm{r})}.
\label{satellites}\end{equation}

The density perturbation, $\delta N$, is found, as in \S\ref{subsec:InducedScattering}, from the Vlasov equation (\ref{eq:Vlasov_lin}) with the ponderomotive potential
\begin{align}
    \varphi
& =\frac 14m\vert a\vert^2={\const}+\frac{ma_0}4(\delta a+\delta a^*)\nonumber\\
& =\const+\frac{ma_0}4 (a_++a_-)e^{i(-\omega_dt+\bm{k}_d\cdot\bm{r})}+{\rm c.c.}
\end{align}
This yields
\begin{equation}
    \frac{\delta N}{N_0}=\frac{a_0m}4(a_++a_-)e^{i(\bm{k}_d\cdot\bm{r}-\omega_dt)}
    \int\frac{k_dv \,d f_0/d\epsilon}{k_dv-\omega_d}\,dp+{\rm c.c.}
\label{dens}\end{equation}
Plugging Eqs.\  (\ref{satellites}) and (\ref{dens}) into the wave equation (\ref{wave_eqn}) and collecting terms with phases $\exp\left[ i(-\omega_dt+\bm{k}_d\cdot\bm{r})\right]$ and $\exp\left[ i(\omega_d^*t-\bm{k}_d\cdot\bm{r})\right]$, one obtains two coupled algebraic equations for $a_+$ and $a_-$,
\begin{align}
    & \left[\pm({\bm v}_g\cdot{\bm k}_d-\omega_d)+\frac{k_d^2-\omega_d^2}{2\omega_0}\right]a_{\pm}\\
    &=\frac{\omega_p^2a_0^2}{8\omega_0}\left(a_+ +a_-\right)
    \left(1
    -m\int\frac{k_dv\, d f_0/d\epsilon} {k_dv-\omega_d}\,dp\right)\nonumber  \, .
\end{align}
The condition that $a_\pm$ have non-trivial solutions gives the dispersion equation
\begin{equation}
\!\!\left[\frac{2\omega_0({\bm v}_g\cdot{\bm k}_d-\omega_d)}{k_d^2-\omega_d^2}\right]^2\!\!=1
-\frac{a_0^2\omega_p^2/2}{k_d^2-\omega_d^2} \left(\!1\!-\!\int\frac{k_dp\,d f_0/d\epsilon}{k_dv-\omega_d}dp\right).
\label{eq:disp1}
\end{equation}

Generally, this equation is solved numerically. We can however obtain analytical results in certain asymptotic regimes. Taking into account that $\omega_0\gg\{\omega_d,\,k_d,\,\omega_p\}$, one sees that
the LHS becomes too large unless $\omega_d\approx{\bm v}_g\cdot{\bm k}_d$. Let us first consider the case of the filamentation instability, where ${\bm v}_g\cdot{\bm k}_d=0$. Then one can neglect $\omega_d^2$ with respect to $k_d^2$. In the integral, one can drop $\omega_d$ only under a more restrictive condition,
\begin{equation}
   \vert \omega_d\vert\ll k_dv_T\, .
\label{eq:omega_d<<k_dv_T}\end{equation}
The condition (\ref{eq:omega_d<<k_dv_T}) ensures that the characteristic instability time is small as compared with the filament crossing time by an electron. In this case, the filaments grow quasistatically: the ponderomotive force is balanced by the plasma pressure at each moment. In the opposite case, which will be considered below, the ponderomotive force is balanced by the electron inertia; then the growth rate of the instability is independent of the plasma temperature.

We first assume that condition \eqref{eq:omega_d<<k_dv_T} is fulfilled (the results will be verified as consistent a posteriori), and later explore the regime where the opposite condition prevails. When $\vert \omega_d\vert\ll k_dv_T$,
\begin{equation}
    \int\frac{k_dp\,d f_0/d\epsilon }{k_dv-\omega_d}dp\simeq m\int\frac{\partial f_0}{\partial\epsilon}dp
    =-\frac m{T}.
\end{equation}
Inasmuch as the plasma is nonrelativistic, $T\ll m$,  the first term in brackets in the RHS of Eq.\ (\ref{eq:disp1})  may be neglected. This term originates from the nonlinear current in the wave equation (the RHS of Eq.\ (\ref{eq:nl_wave_eq}) and the last term in Eq.\ (\ref{wave_eqn})), so the nonlinear current does not play a role in the filamentation instability of the pair plasma; the effect is determined only by the ponderomotive force.  The dispersion equation now reduces to
\begin{equation}
    \frac{4\omega_d^2\omega_0^2}{k_d^2}=
    k_d^2-\frac{m}{2T}\omega_p^2a^2_0 \, .
    \label{eq:displow}
\end{equation}
Hence, the instability develops if $k_d<(m/2T)^{1/2}\omega_pa_0$
, as $\omega_d$ becomes purely imaginary. The maximum growth rate,
\begin{equation} \label{eq:FilamentationGrowthRate}
    \Gammamax={\rm Im}\,\omega_{d,\rm max}=\frac m{8T}\frac{\omega_p^2}{\omega_0}a^2_0\,,
\end{equation}
is achieved at
\begin{equation}
\label{eq:FilamentationFastestMode}
 k_{d}=\frac 12\left(\frac mT\right)^{1/2}\omega_pa_0\,.
\end{equation}
One sees that the growth rate of the filamentary instability is of the order of the induced scattering rate (\ref{kappa_scat_max}).
This result was obtained under the condition (\ref{eq:omega_d<<k_dv_T}), i.e., for $\Gamma={\rm Im}\,\omega_d\ll k_dv_T$.
Making use of Eqs.\ (\ref{eq:FilamentationGrowthRate}) and (\ref{eq:FilamentationFastestMode}), the result is seen to be applicable for
\begin{equation}
    \frac{\omega_pa_0}{\omega_0}\ll\frac{4T}{m}.
\label{eq:conditionFilament}\end{equation}

\begin{figure}[h!]
    \centering
    \includegraphics[width=0.49\textwidth]{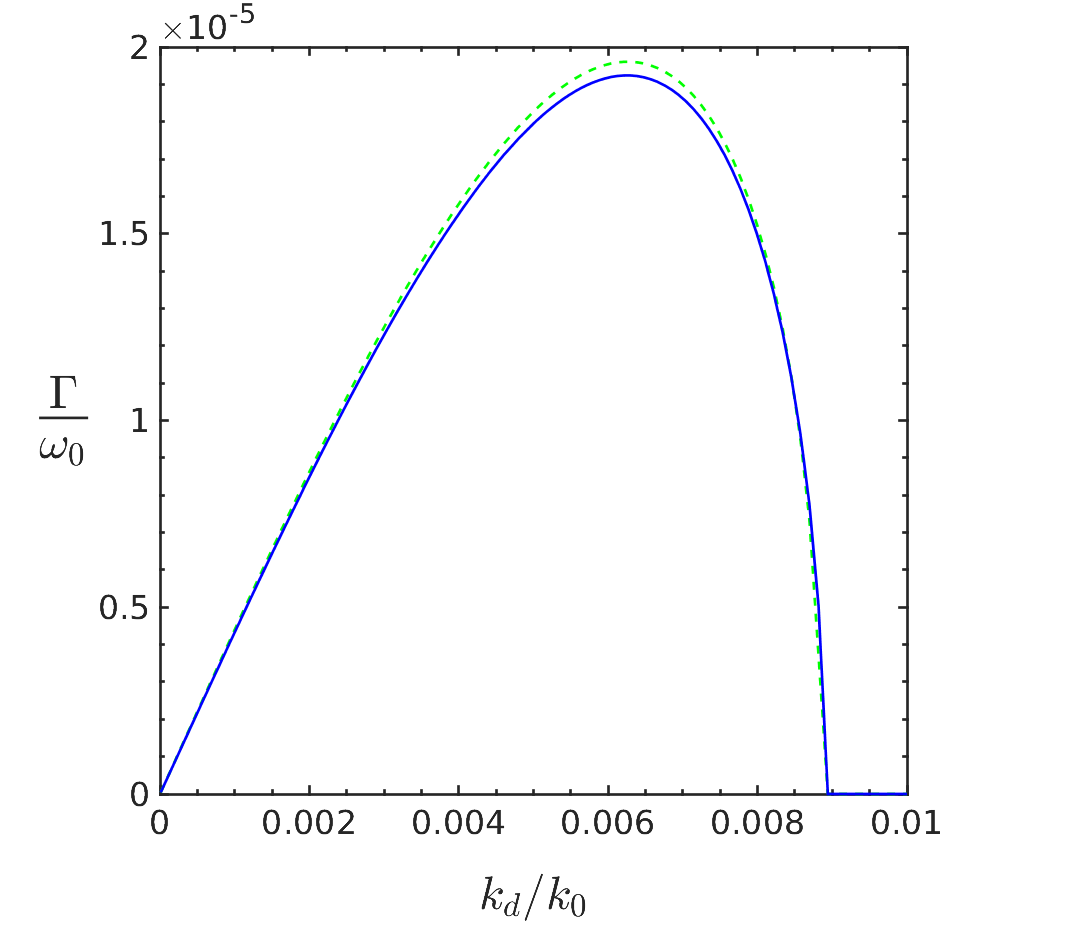}
    \caption{The growth rate of the filamentation instability, 
    normalized to the incident angular frequency, evaluated by numerical integration of the dispersion relation \eqref{eq:disp1} (solid blue) and according to the asymptotic theoretical approximation \eqref{eq:displow} (dashed green), which is valid under the condition \eqref{eq:conditionFilament}. Parameters: $T/m = a = 0.00137$, $\omega_p/\omega_0 = 0.107$, and a 1D Maxwellian $f_0$.
    \label{fig:numfila}}
\end{figure}

In Fig.~\ref{fig:numfila}, we present the growth rate of the filamentation instability found both from the numerical solution to the dispersion  equation (\ref{eq:disp1}) and from the asymptotic formula (\ref{eq:displow}). For this purpose, we use parameters for which the condition (\ref{eq:conditionFilament}) for the validity of the asymptotic solution is satisfied. Solutions for a wide range of parameters are presented farther below as we compare the results of PIC simulations of the filamentation instability with the theory.

In the limit $k_d v_T\ll|\omega_d|\ll k_d$, one can expand the integral on the RHS of Eq.\ (\ref{eq:disp1}) in small $k_dv$:
\begin{equation}
    \int\frac{k_dp\,df_0/d\epsilon}{k_dv-\omega_d}dp \simeq -\frac{k_d^2}{\omega_d^2}\int p\frac{df_0}{dp}dp=\frac{k_d^2}{\omega_d^2}.
\end{equation}
As $k_d\gg|\omega_d|$, one can again neglect the first term in brackets on the RHS of Eq.\ (\ref{eq:disp1}).
Hence, the nonlinear current does not play a role, whether or not condition (\ref{eq:omega_d<<k_dv_T}) is satisfied.
In the present case, the dispersion equation takes the form
\begin{equation}
    \frac{4\omega_d^2\omega_0^2}{k_d^4}=1+\frac{\omega_p^2a^2_0}{2\omega_d^2}\,. \label{eq:disphigh}
\end{equation}
This bi-quadratic equation always has an unstable solution. The growth rate increases with increasing $k_d$, and reaches
\begin{equation}
    \Gammamax=\omega_pa_0/\sqrt{2}\,,
\label{eq:filamentMaxGrowth1}\end{equation}
at $k_d=\infty$. Expansion in $k_d^{-1}$ yields
\begin{equation}
    \Gamma=\left(1-\frac{2\omega_p^2\omega_0^2a_0^2}{k_d^4}\right)\Gammamax.
\end{equation}
One sees that at
\begin{equation}
k>k_{d,1}= 3(\omega_p\omega_0a_0)^{1/2}\,,
\end{equation}
the growth rate approaches the maximal one, Eq.\ (\ref{eq:filamentMaxGrowth1}), to within 2.5\%. On the other hand,
the condition $\Gamma\gg k_dv_T$ is violated at
\begin{equation}
    k>k_{d,2}\simeq a_0\omega_p(m/2T)^{1/2}>k_{d,1} \,.
\label{eq:filamentMaxk1}\end{equation}
This in fact means, that all perturbations with the wave vectors $k_{d,1}<k_d<k_{d,2}$ grow with practically the same growth rate (\ref{eq:filamentMaxGrowth1}).

Now let us turn to the case of modulation at some angle $\theta_d$ with respect to the pumping wave. We may again approximate $\omega_d\simeq\bm{k}_d\cdot\bm{v}_g$ in the dispersion equation (\ref{eq:disp1}), otherwise its LHS becomes too large. When $\theta$ is sufficiently far from $\pi/2$, such that $\mu_d\equiv\cos\theta_d\gg \sqrt{T/m}$, one can then expand the integral in the RHS of the equation in small $v$,
\begin{equation}
    \int\frac{p\,df_0/d\epsilon}{v-v_g\mu_d}dp \simeq
    -\frac 1{v_g^2\mu_d^2}\int p\frac{d f_0}{dp}dp=\frac 1{v_g^2\mu_d^2}\,.
\end{equation}
Then the dispersion equation becomes
\begin{equation}
    \left[\frac{2\omega_0({\bm v}_g\cdot{\bm k}_d-\omega_d)}{k_d^2(1-v_g^2\mu_d^2)}\right]^2=1
    +\frac{1}{2}\left(\frac{\omega_p a_0}{k_d v_g \mu_d}\right)^2\,.
\end{equation}
The RHS of the dispersion equation is seen to be positive, implying that the instability does not develop in this case.

One thus concludes that in a pair plasma, only the filamentation instability develops, so the pumping wave is modulated in the direction perpendicular to $k_0$. The main effect in this case is density variations due to the ponderomotive force. In an electron-ion plasma, the density variations are suppressed because of a high ion inertia. In this case, the effect of the nonlinear current comes into play; then both the filamentation and the modulation instabilities could develop \citep{Max_etal74,Sobacchi_etal20}.

An important point is that the ponderomotive force depends only weakly on the radiation spectrum provided it is composed from waves with random phases. In this case, one substitutes
$\bm{A}^2$ in Eq.\ (\ref{ponderomotive})
by $\int\vert\bm{A}_{\omega}\vert^2d\omega$. The variations of the plasma density, which lead to the focusing of the radiation, are produced by the total ponderomotive force, therefore the filamentation instability is determined by the total radiation energy and the characteristic frequency, the growth rate of the instability being only weakly dependent on the shape of the radiation spectrum.

Formally, one has to substitute the ansatz (\ref{eq:complex_A}) by
\begin{equation}
   \frac em A={\rm Re}\,\sum_{\omega_0}\left[1+\delta a(\bm{r},t)\right]a_{\omega_0}e^{-i(\omega_0t-k_0x)},
\end{equation}
where $\vert a_{\omega_0}\vert^2$ is the spectral distribution of the initial beam, $\delta a(\bm{r},t)\ll 1$ describes the common envelope. As we have already seen, the nonlinear current (\ref{eq:nl_current}) may be neglected when considering the filamentation instability in the pair plasma. This means that one can neglect the $a^2_0$ term in the dispersion equation for the pumping wave, Eq.\ (\ref{eq:plane_wave}) and also  neglect the term in the RHS of the wave equation (\ref{eq:nl_wave_eq}).  Substituting the new anzats in thus modified wave equation and linearizing with respect to small $\delta a$ and $\delta N/N_0$, we get (cf.\ Eq.\ (\ref{wave_eqn}))
\begin{equation}
2i\omega_0\left(\frac{\partial \delta a}{\partial t}+v_g\frac{\partial \delta a}{\partial
x}\right)-\frac{\partial^2 \delta a}{\partial
t^2}+\Delta\delta a
-\omega_p^2\frac{\delta
N}{N_0}=0.
\label{eq:modif_wave_eq}\end{equation}
The ponderomotive potential is now written as
\begin{align}
    \varphi
& =\frac 14m\vert (1+\delta a)^2\vert^2\int\vert a_{\omega}\vert^2d\omega\\
& =\const+\frac{m}4 (a_++a_-)e^{i(-\omega_dt+\bm{k}_d\cdot\bm{r})}\int\vert a_{\omega}\vert^2d\omega+{\rm c.c.},\nonumber
\end{align}
which yields \ Eq.\ (\ref{dens}), in which $a_0^2$ should be substituted by $\int\vert a_{\omega}\vert^2d\omega$.
An important point is that in the case under consideration, $\omega_d$ and $\bm{k}_d$ are independent of $\omega$ because the envelope is common for the whole pumping beam.
Substituting thus obtained $\delta N$ into Eq.\ (\ref{eq:modif_wave_eq}), one finds the dispersion equation similar to (\ref{eq:disp1}), but without the first term in the brackets  in the RHS (which is not important anyway) and with $a_0^2$ substituted by $\int\vert a_{\omega}\vert^2d\omega$. Therefore the same substitution should be done in all the obtained results for the filamentary instability.

As shown in \S\ref{subsec:InducedScattering}, the induced scattering rate decreases with the width of the spectrum. Therefore, although the growth rates of induced scattering and the filamentation instability are comparable to each other for a monochromatic wave, in the case of a wide radiation spectrum, the filamentation instability develops significantly faster than induced scattering. This conclusion is valid only for a pair plasma, because in an electron-ion plasma, the growth rate of the filamentation instability is diminished \citep{Max_etal74,Sobacchi_etal20}.

\section{PIC simulation setup}
\label{sec:simulations}

We carry out a suite of PIC simulations to investigate the non-linear interactions of radiation with pair plasma. The primary code used in this paper is EPOCH \citep[][version 4.17.10; \href{link}{https://github.com/Warwick-Plasma/epoch}]{Arber_2015}, while some simulations are also carried out using the code Tristan-MP \citep[][version 1; \href{link}{https://github.com/ntoles/tristan-mp-pitp}]{spitkovsky_structure_2008}, for the purposes of comparison and verification. We focus on EPOCH because of the numerical considerations discussed in \S\ref{sec:numstrong}, and for the efficiency gains resulting from the 1D simulation option it provides. Here, we outline the setup of the simulations and their numerical considerations, methods, and convergence, postponing the results to \S\ref{sec:results}.

\subsection{Initial conditions}
\label{subsec:InintialConditions}

We are interested in the time evolution of the pair plasma due to its interaction with a strong electromagnetic wave.
Therefore, we initialize the simulations with a homogeneous, neutral, thermal plasma distribution, and introduce a planar, monochromatic wave, called the pumping wave.
The wave has the form
\begin{equation}
\bm{E}=E_0 \sin\left(k_0 x\right) \hat{\bm{y}}\, ; \quad
\bm{B}=E_0 \sin\left(k_0 x\right) \hat{\bm{z}} \, ,
\end{equation}
as illustrated in Fig.~\ref{fig:illust}, filling the simulation box.
Here, $k_0=2\pi/\lambda_0$ and $\lambda_0$ are the angular wavenumber and wavelength.
The amplitude $E_0=a m \omega_0/\mathrm{e}$ of the electric and magnetic fields is defined in accordance with Eq.~(\ref{eq:a}), where $\omega_0=k_0$ is the angular frequency of the wave in vacuum.

We explore 2D setups for the study of both induced scattering and the filamentation instability.
Taking advantage of the preferential backscattering of the former, we utilize 1D setups to explore induced scattering in more detail. Periodic boundary conditions are applied in all directions of the simulation box, as applicable in each case. The particles are initialized in a relativistic Maxwellian, accounting for oscillations in the field of the wave, as discussed in \S\ref{sec:drifts}.

\subsection{Electromagnetic wave initialization}
\label{sec:numstrong}

Initializing a strong wave in a PIC code involves several subtle considerations. First, the electric and magnetic fields are located in different locations within the Yee lattice, and are advanced at different times. This combination greatly reduces the discretization errors, but require that the initial electric and magnetic fields be phase-shifted relative to each other, by a distance of the order of the grid length, in order to correctly capture the physical behavior of the electromagnetic wave. The initial magnetic field (on its staggered grid) thus requires an added phase (with respect to the initial electric field) of $\Delta x+2\Delta t$ in EPOCH and $\Delta x+\Delta t$ in Tristan-MP, where $\Delta x$ is the grid spacing and $\Delta t$ is the time step.

Second, there are errors associated with the interpolation of the fields to the particle locations, which is needed to calculate the electromagnetic force in the particle mover. In the Tristan-MP version used in the analysis, only linear interpolation is available, which entails a significant fractional error in the electromagnetic field, of order
\begin{equation} 1-\frac{\sin[k_0(x+\frac{\Delta x}{2})]+\sin[k_0(x-\frac{\Delta x}{2})]}{2\sin (k_0 x)}\simeq \frac{(k_0 \Delta x)^2}{8}\, .
\end{equation}
We explicitly alter the interpolation routine to correct the results specifically for the case of a monochromatic pumping wave. With this correction, we are able to demonstrate good agreement between EPOCH and Tristan-MP in this case (see Appendix \S\ref{app:Convergence}). For multiple modes, such a tailored correction is no longer possible with linear interpolation, so we focus on EPOCH, which uses a triangular (second order) interpolation.

Third, the discrete nature of the PIC code changes the speed of light. This effect exists for both EPOCH and Tristan-MP. Fourth-order interpolation in the field solver, employed in both codes,
greatly reduces this source of error. For a sufficiently long wavelength relative to the grid spacing, the exact speed of light does not have a strong effect on the dynamics. One should however take the corrected speed of light into account, for example when using test particles.

Convergence tests, which indirectly also gauge these three problems, are discussed in \S\ref{subsec:Convergence}.

\subsection{Particle initialization and orbit drifts}
\label{sec:drifts}

As our simulations involve strong waves, initializing the particles in a thermal distribution with no regard to the field would lead to a violent energy conversion at the very beginning of the simulation. To reduce this effect, we place each particle in an orbit in the field of the vacuum electromagnetic wave (where the angular frequency is $\omega_0$, and a precise solution for the particle orbits is known), as follows.

The field of the wave is given in this case by the vector potential
\begin{equation}
    \bm{A}=\bm{\hat{y}}\frac{ E_0}{\omega_0}\cos(\eta) \, ,
\end{equation}
where
\begin{equation}
\eta = \omega_0 (t-x)
\end{equation}
is the phase. The particle orbit is described by the first integrals of motion (e.g., \citealt{Melrose_book80}), written in the notations of \citet{Lyubarsky19a} as
\begin{equation}
    w=u_y+a\cos\eta
\end{equation}
and
\begin{equation}
    g= \gamma-u_x\, ,
\end{equation}
where $u_x$ and $u_y$ are the respective components of the particle 4-velocity $\bm{u}=\gamma\bm{v}$, and $\gamma=(1-v^2)^{-1/2}$ is the Lorentz factor. The particle velocities are expressed via the integrals of motion as
\begin{gather}
 v_x=\frac{1+(w-a\cos\eta)^2-g^2}{1+(w-a\cos\eta)^2+g^2};\label{eq:betax}\\
v_y=\frac{2g(w-a\cos\eta)}{1+(w-a\cos\eta)^2+g^2}.
\end{gather}
The velocity of the particle guiding center (denoted by tilde) is found by averaging the velocities over the wave period, $\tilde{\bm{v}}=\langle\bm{v}\rangle$.
The integrals of motion are expressed via the velocities of the particle guiding center,
\begin{gather}
    g=(1+a^2/2)^{1/2}\tilde{\gamma}(1- \tilde{v}_{x})\,;\\
    w=(1+a^2/2)^{1/2}\tilde{\gamma}\tilde{v}_y\, , \label{eq:weq}
\end{gather}
where $\tilde{\gamma}=(1-\tilde{v}^2)^{-1/2}$ is the Lorentz factor of the particle guiding center.

To initialize our particles in approximate orbits, we first calculate the guiding center velocity, $\tilde{\bm{v}}$, and the corresponding Lorentz factor from a relativistic Maxwellian at the initial temperature $T$. Then we calculate the phase, $\eta=-\omega_0 x_0$, at the initial position $x=x_0$ and time $t=0$. Finally, we use Eqs.~(\ref{eq:betax})--(\ref{eq:weq}) to calculate the initial values of $v_x$ and $v_y$ for the particle. This procedure results in much milder numerical transients.

\subsection{Measuring growth rates in simulations}
\label{subsec:GrowthRates}

Sufficiently large simulations in 2D can resolve both induced scattering and the filamentation instability, simultaneously.
However, to better study each process separately, we focus on two types of simulations:
1D simulations along the pumping wave, which can only resolve induced scattering, and 2D simulations confined to only one wavelength along the beam but many wavelengths in the perpendicular direction, which can only resolve the filamentation instability.

Both induced scattering and the filamentation instability result in perturbations in both the density and the electromagnetic field.
We measure these perturbations in our simulations, in each diagnosed time step (snapshot), as follows.
For the density field, we compute the root mean square (rms) of fractional density perturbations $\delta N_{\rm rms}/N_0$  across the entire simulation box. For the electromagnetic perturbations, we perform a Fourier decomposition in traveling modes of wavenumber $\bm{k}$, as described in Appendix \ref{sec:expansion}, and compute the normalized spectral distribution of the Poynting flux, $\bm{S}_{\bm{k}}/S_0$, where $S_0\equiv S(t=0;\bm{k}=\bm{k}_0)$ is the initial beam flux.
We also measure the normalized total electromagnetic energy outside the beam,
\begin{equation}
\label{eq:PICScatteredFluxgeneral}
\frac{\delta  S}{S_0}
= \frac{\sum_{\bm{k}\neq \bm{k}_0}{\langle S_{\bm k}\rangle}}{S(t=0;\bm{k}=\bm{k}_0)} \, ,
\end{equation}
where $t=0$ corresponds to the beginning of the simulation.
More generally, when injecting multiple modes, $\delta S/S_0$ is defined as the flux in all other modes, normalized to the initial, injected flux.

The snapshots of a simulation are labelled by an integer $\tau=0,1,2,\ldots, \tau_{\mathrm{max}}$. The growth rate of a given quantity is measured by fitting an optimal range of snapshots during an exponential growth phase.
The time interval between snapshots is denoted $\Delta \mathbb{T}$ and chosen such that $1 \lesssim \Delta \mathbb{T}/\lambda_0 \lesssim 2\pi T/(\omega_p^2 \lambda_0^2 m a^2)$, in order to resolve the theoretical growth rate (of both induced scattering and filamentation), but not necessarily the pumping wave frequency. Our choice of $\Delta \mathbb{T}/\lambda_0$ varies from $1$ to $100$, depending on the simulation parameters.

To outline the general procedure to measure growth rates from simulations, we denote the relevant perturbed quantity as $Q$; namely, $Q=\delta S/S_0$ for 1D (induced scattering) simulations, and $Q=\delta N_\mathrm{rms}/N_0$ for 2D (filamentation) simulations.
To minimize noise and transients effects in the growth-rate measurement, we exclude early-time snapshots where $Q$ is small.
To avoid saturation effects, we exclude late-time snapshots where $Q$ is large. Quantitatively,
our simulations indicate that the growth of $Q$ 
may be susceptible to these spurious effects outside the interval $10^{-3}<\delta{S}/S_0<10^{-2}$ in 1D, and the interval $10^{-2}<\delta N_\mathrm{rms}/N_0<2.5\times 10^{-2}$ in 2D.

Consider two snapshots, $\tau_1$ and $\tau_2$, confined within the relevant interval above.
Modeling $Q$ as exponential growth during all snapshots between $\tau_1$ and $\tau_2$ yields a best fit $d(\ln Q)/dt=\mu\pm\sigma$. The values of $\tau_1$ and $\tau_2$ are selected as the cleanest exponential behavior, identified with the minimal $\sigma$. Finally, depending on $Q$, we assign the measurement $\mu\pm \sigma$ to either $\kappa$ or $\Gamma$.

\subsection{Simulation parameters and convergence}
\label{subsec:Convergence}

The nominal physical parameters in our simulations are chosen, as in Figs.~\ref{fig:numscat} and \ref{fig:numfila}, to represent a cold plasma with $T/m = 0.00137$ and a monochromatic pumping wave with a short wavelength $\omega_p/k_0=0.107$.
The pumping wave amplitude $a = 0.034$ is chosen larger than in the above theoretical figures, in order to accelerate perturbation growth.
The nominal numerical parameters are chosen differently for 1D and for 2D simulations, otherwise the latter would require too many resources.

For 1D simulations, we take $N_{ppc}=120$ particles per cell (both species) and $\lambda_0/\Delta x=160$ cells per wavelength, \ie $(\omega_p \Delta x)^{-1}=238$ cells per skin depth.
The simulation box length is chosen as $L_x/\lambda_0=200$ wavelengths, sufficient to spectrally resolve induced scattering.
To see this, note that the wavelength shift for the fastest growing mode, as given by Eq. \eqref{eq:Backscattering}, is very small ($\lesssim 10\%$ for our nominal value of $T/m = 0.00137$), so the box must be larger than
\begin{equation}
    \frac{2\pi}{|\bm{k_0}| - |\bm{k_1}|} \simeq \frac{1}{2\sqrt{T/m}}\, ,
\end{equation}
giving $\sim 14 \lambda_0$ for nominal physical parameters.
Our Tristan-MP simulations are effectively 1D, using only one non-guard cell in the $y$ direction.

The nominal numerical parameters for the 2D simulations are chosen as $N_{ppc}=20$ particles per cell and $\lambda_0/\Delta x= 80$ cells per wavelength, \ie $(\omega_p \Delta x)^{-1}=119$ cells per skin depth.
The simulation box consists of square, $\Delta x=\Delta y$ cells, and has dimensions $L_x/\lambda_0 = 1$, $x$ being the incident wave direction, and $L_y/\lambda_0=100$.
The extent along the $y$ direction is chosen to accommodate the typical wavelength of filamentation in Eq.\ (\ref{eq:FilamentationFastestMode}),
\begin{equation}
\frac{\lambda_f}{\lambda_0} \simeq \frac{2\omega_0}{\omega_p a}\sqrt{\frac Tm}\, ,
\end{equation}
giving $\lambda_{f} \sim 20 \lambda_0$ for nominal physical parameters.

We carry out a suite of convergence tests, and find good convergence (to $15\%$ in growth rate) in each of our numerical parameters $n_\mathrm{ppc},\text{ }\lambda_0/\Delta x \text{ and } L/\lambda_0$. The agreement between EPOCH and Tristan-MP and the parametric convergence are demonstrated in Appendix \ref{app:Convergence}.

Simulation times are normalized to the angular frequency $\omega_0$ of the pumping wave in vacuum.
In addition, for easier comparison with theory, we also show times normalized to the theoretical maximal growth rates. For our nominal physical parameters, used in all evolution figures below, Eq.~\eqref{eq:scatdisp} yields $\kappa_\mathrm{max} = 1.94 \times 10^{-3} \omega_0$ for induced scattering, and Eq. \eqref{eq:disp1} yields $\Gamma_\mathrm{max} = 6.98 \times 10^{-4} \omega_0$ for filamentation.

For brevity, nominal 1D simulations henceforth refer to 1D (induced scattering) EPOCH simulations with a monochromatic pumping wave and nominal physical and numerical parameters.
Similarly, nominal 2D simulations refer to 2D (filamentation) EPOCH simulations with a monochromatic wave and nominal parameters.

\section{Results}
\label{sec:results}

We present the results from our 1D simulations, which resolve only  induced scattering, in  \S\ref{subsec:ISResults}. Results from our 2D simulations, which resolve only the  filamentation instability, are presented in \S\ref{subsec:FilamentationResults}.
Overall, the results demonstrate a good agreement with the theory throughout the relevant parameter and dynamical ranges.

\subsection{Induced scattering}
\label{subsec:ISResults}

Here we study the physical manifestations of induced scattering, from onset to saturation, and compare them to the theory.
The case of a monochromatic wave is discussed in \S \ref{subsub:Monochromatic}.
The more realistic scenario, of an incident wave with a broad spectrum, is studied in \S \ref{subsub:broadspec}. 

\subsubsection{Monochromatic beam}
\label{subsub:Monochromatic}

\begin{figure}[h!]
\centering
\includegraphics[trim={2.5cm 0.5cm 1.5cm 0.5cm},clip,width=0.5\textwidth]{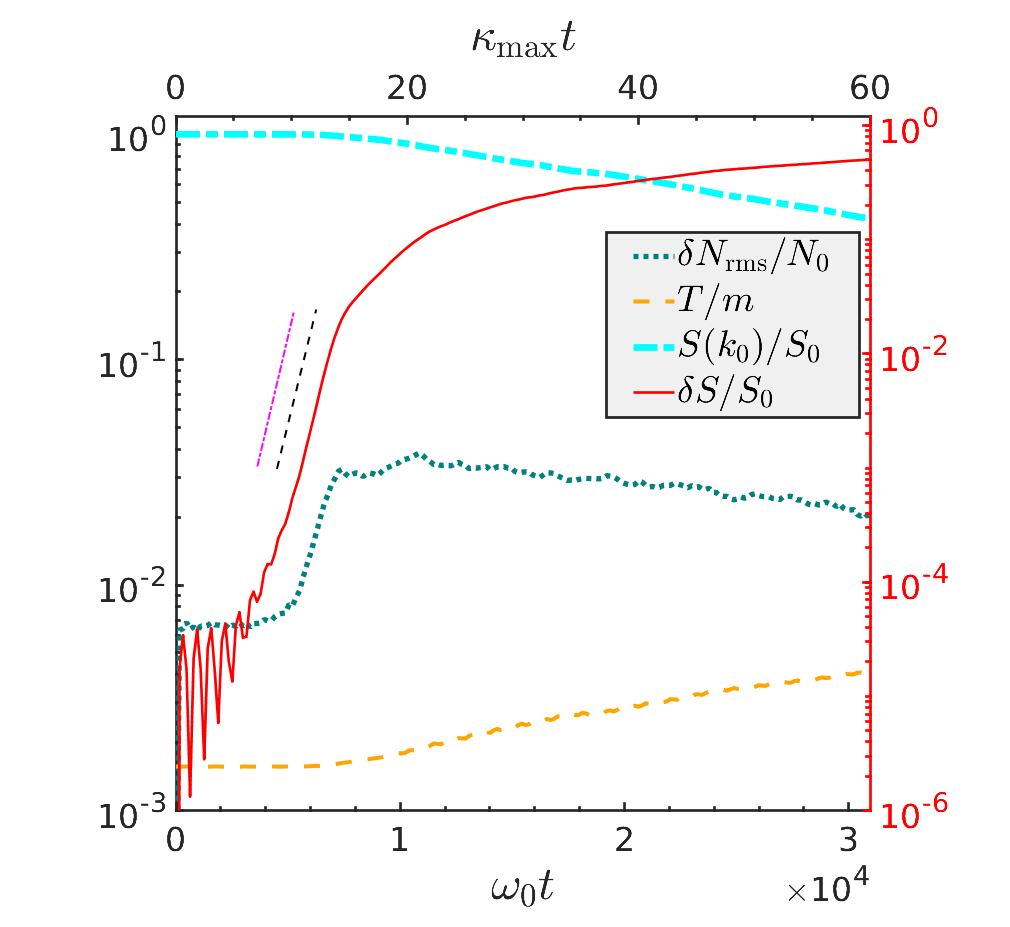}
\caption{Evolution of the nominal pair plasma and strong incident monochromatic wave in a 1D (induced scattering) simulation, showing the normalized evolution of (see legend) the incident beam energy (dot-dashed cyan), scattered electromagnetic flux (solid red with right axis), density perturbations (teal dotted; after removing small-scale noise by applying a high, $k > 4k_0$ filter; see text), and average plasma temperature (orange dashed).
The nominal (physical and numerical parameters, EPOCH) 1D simulation is modified with a high $N_{ppc} = 1200$ (but smaller $L_x = 100\lambda_0$) to resolve the density perturbations. The measured growth rate of $\delta S/S_0$ agrees with its theoretical fastest growth rate (thin magenta dot-dashed slope; obtained by solving Eq.~\eqref{eq:scatdisp} numerically) and our fitting procedure (thin black dashed slope).\label{fig:densgrowth}
}
\end{figure}

\begin{figure}[h!]
\centering
\includegraphics[width=0.48\textwidth]{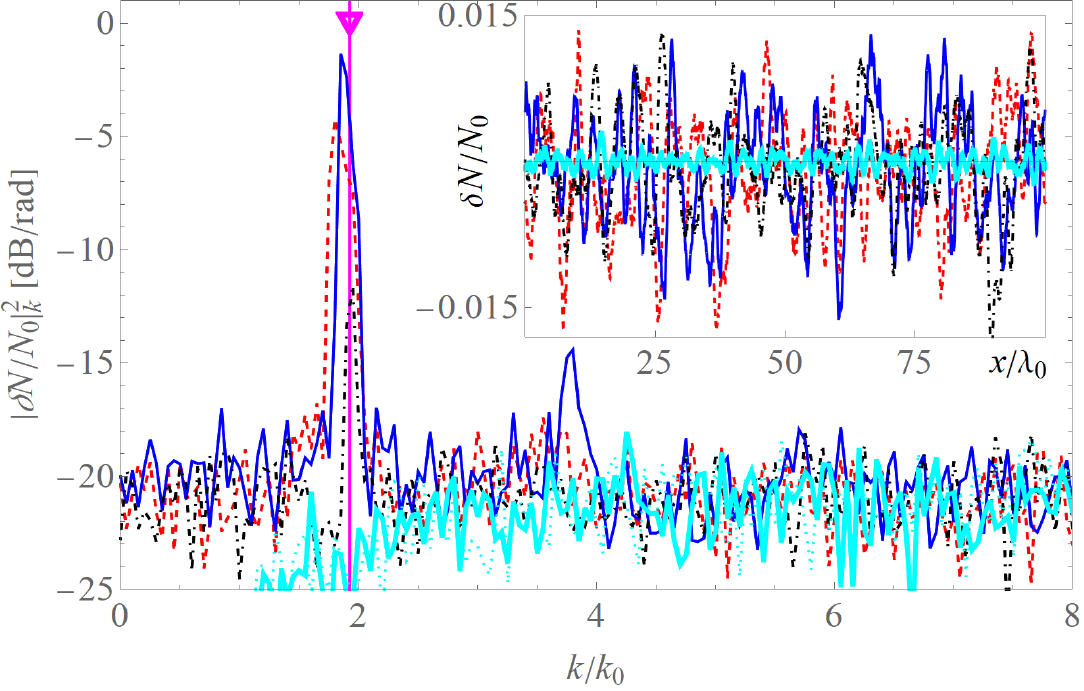}
\caption{
Density perturbations (thin curves) in a nominal 1D (induced scattering) simulation shown during early
($t\simeq3600\omega_0^{-1}\simeq 7\kappamax^{-1}$; dot-dashed black) and late ($t \simeq 7700\omega_0^{-1} \simeq 15\kappamax^{-1}$; solid blue) exponential growth, and in saturation phase ($t \simeq 15000\omega_0^{-1} \simeq 30\kappamax^{-1}$; dashed red).
Beating at an angular wavenumber $k_d$ is seen in the density field (inset; smoothed with a Gaussian of standard deviation $\lambda_0$) and as the main peak in the power spectrum (of data folded over $20\lambda_0$ to reduce noise), consistent with the expected (magenta triangle and vertical line) $k_d/k_0\simeq 2-2(T/m)^{1/2}$ of Eq.~\eqref{eq:Backscattering}.
A shift to slightly lower $k$ and the growth of the first harmonic are seen to develop at the later times.
Fluctuations in electric charge $(N_p-N_e)/N_{\rm tot}$ (thick cyan, for the later time) remain small and consistent with small-scale white noise.\label{fig:periodogram}
}
\end{figure}

\begin{figure*}[t!]
     \centering
    \includegraphics[trim={0cm 0.7cm 2.5cm 0.5cm},clip,width=0.5\textwidth]{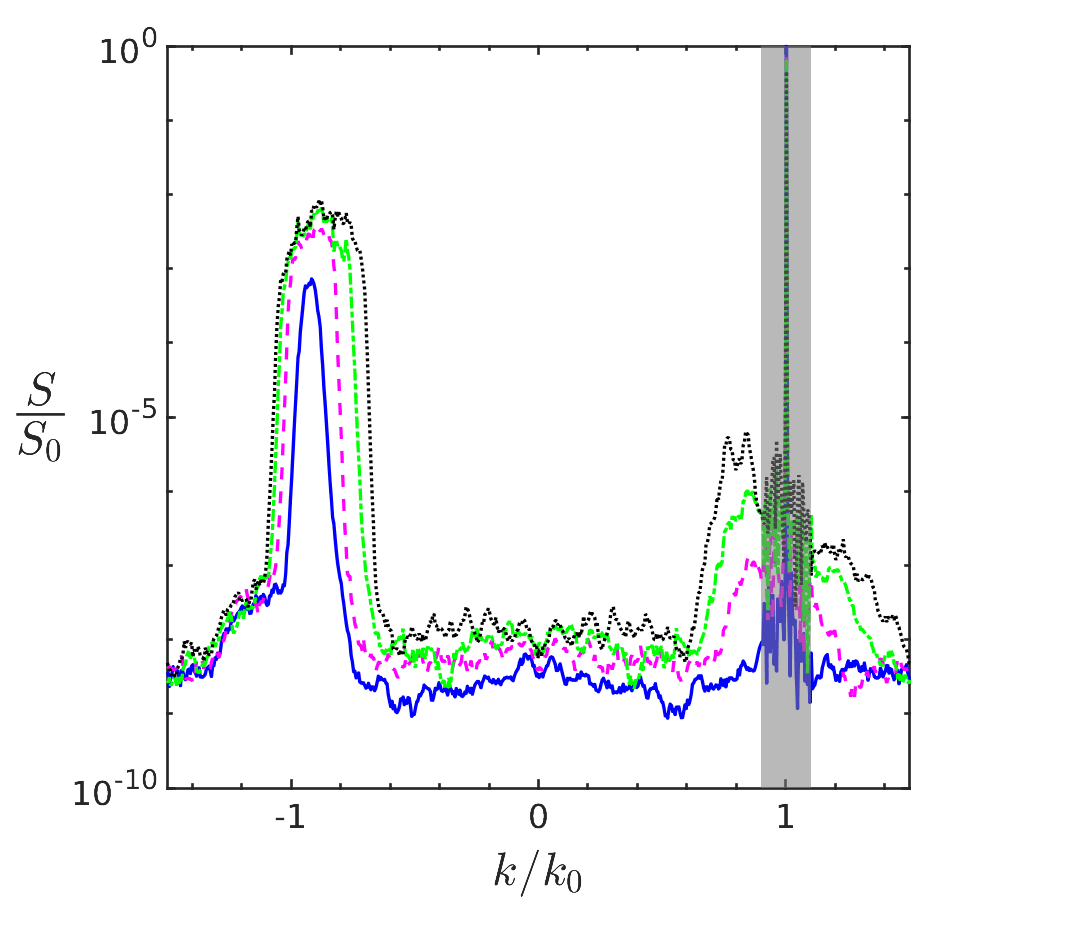}
    \includegraphics[trim={0.5cm 0.5cm 0.5cm 0.5cm},clip,width=0.49\textwidth]{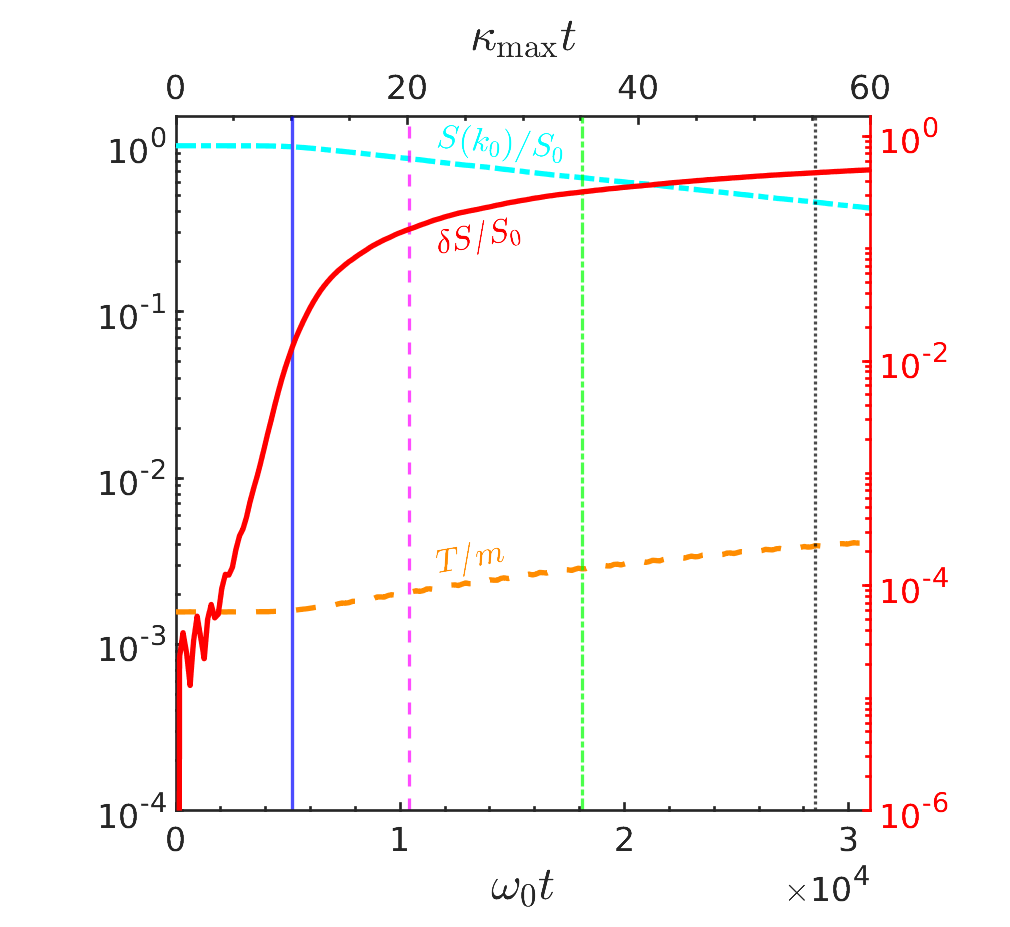}
        \caption{
        Evolution of the electromagnetic spectrum in a nominal 1D (induced scattering) simulation (left panel), with an evolution plot for reference (right panel, notations as in Fig.~\ref{fig:densgrowth}).
        The spectrum of electromagnetic
        flux normalized to the incident
        flux, $S(k)/S_0$, is plotted (curves in left panel, with coincidental vertical lines in the right panel) at times
        $\kappa_\mathrm{max}t \simeq 10$ (solid blue), $20$ (dashed magenta), $35$ (dash-dotted green), and $55$ (dotted black).
        The spectrum has been smoothed
        using a moving average of
        ten $k$-modes,
        except in the shaded region, which
        contains the pumping, 
        $k=k_0$ wave.
        }
        \label{fig:specevo}
\end{figure*}

\begin{figure}[h!]
\begin{center}
\includegraphics[trim={1cm 0.8cm 2cm 0cm},clip,width=0.5\textwidth]{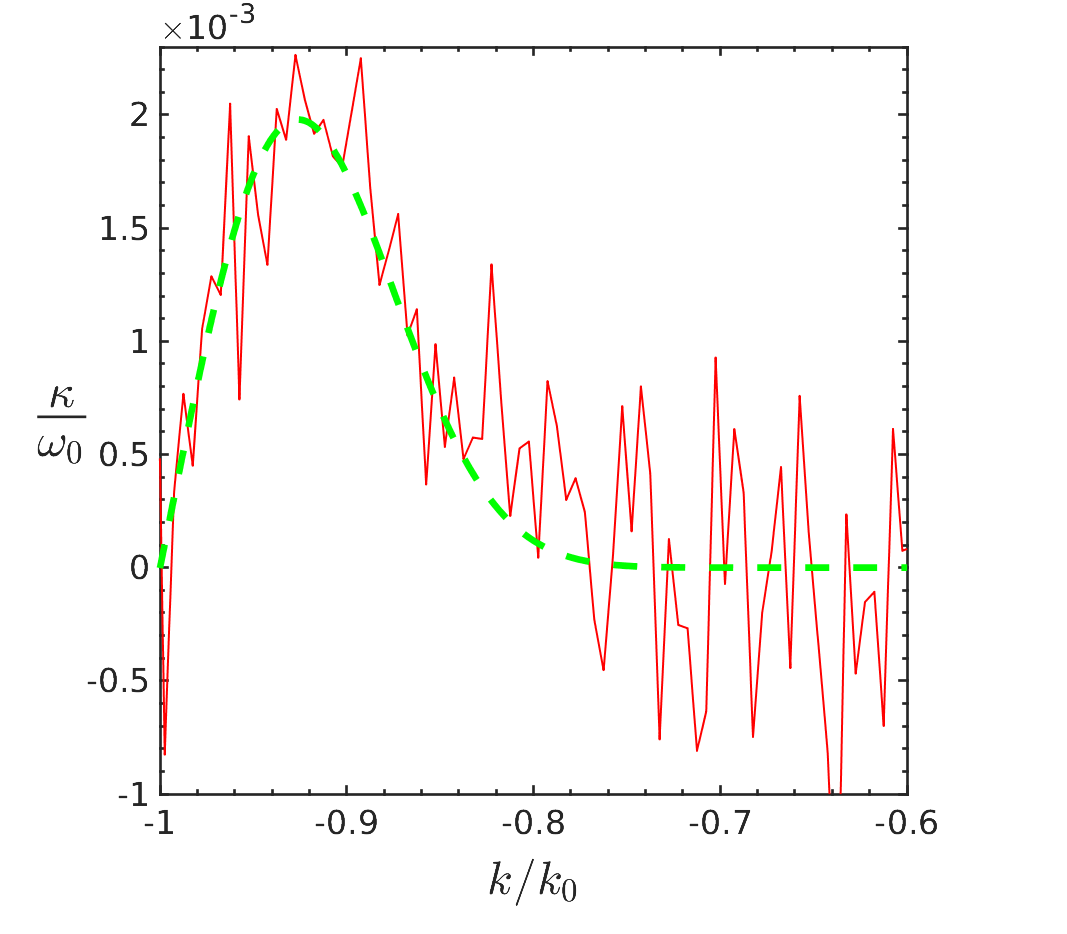}
\end{center}
\caption{
The mean growth rate of different electromagnetic $k$-modes in a nominal 1D (induced scattering) simulation during the exponential growth phase (time interval $[t_1,t_2]=[4,8]\kappa_\mathrm{max}^{-1}$; solid red), compared to the asymptotic theory (Eq.~\eqref{eq:finalgrowth}; dashed green).
\label{fig:dfcomp}}
\end{figure}

\begin{figure*}[t!]
     \centering
     \includegraphics[width=0.48\textwidth]{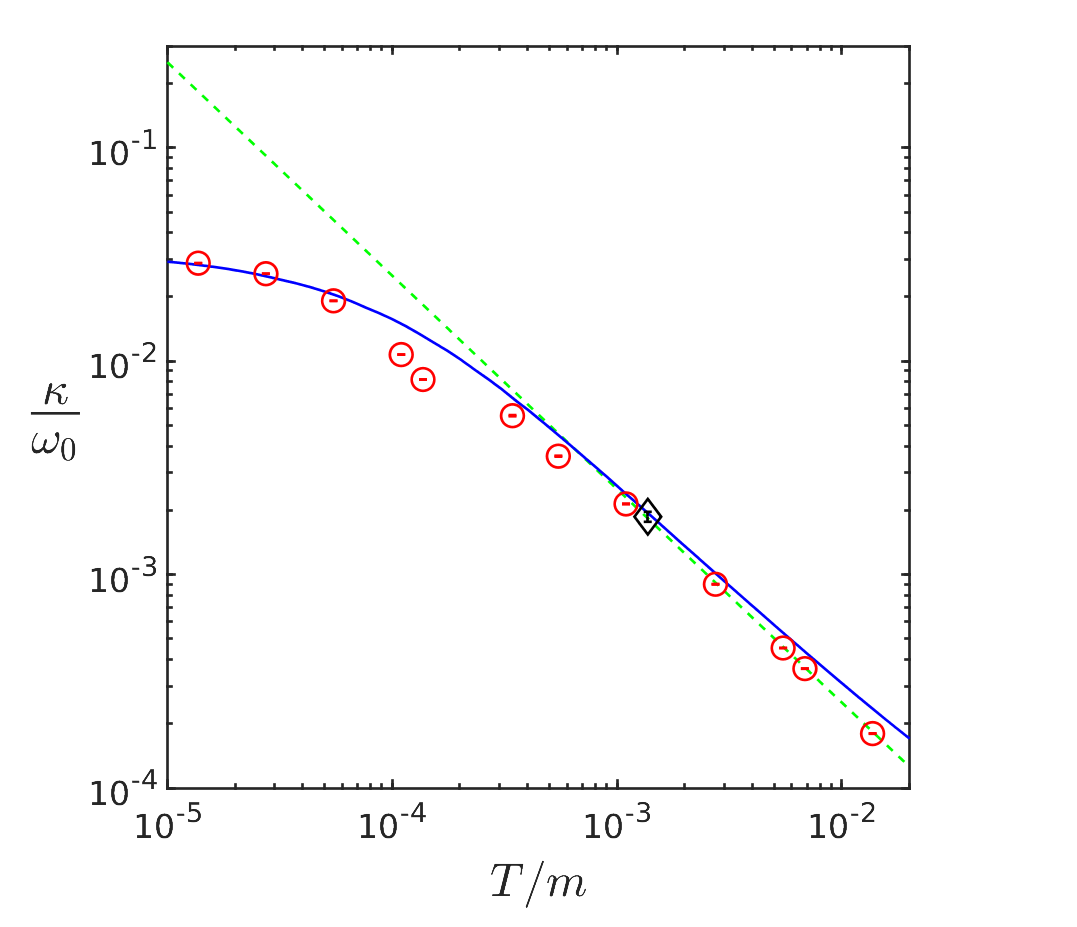}
     \includegraphics[width=0.48\textwidth]{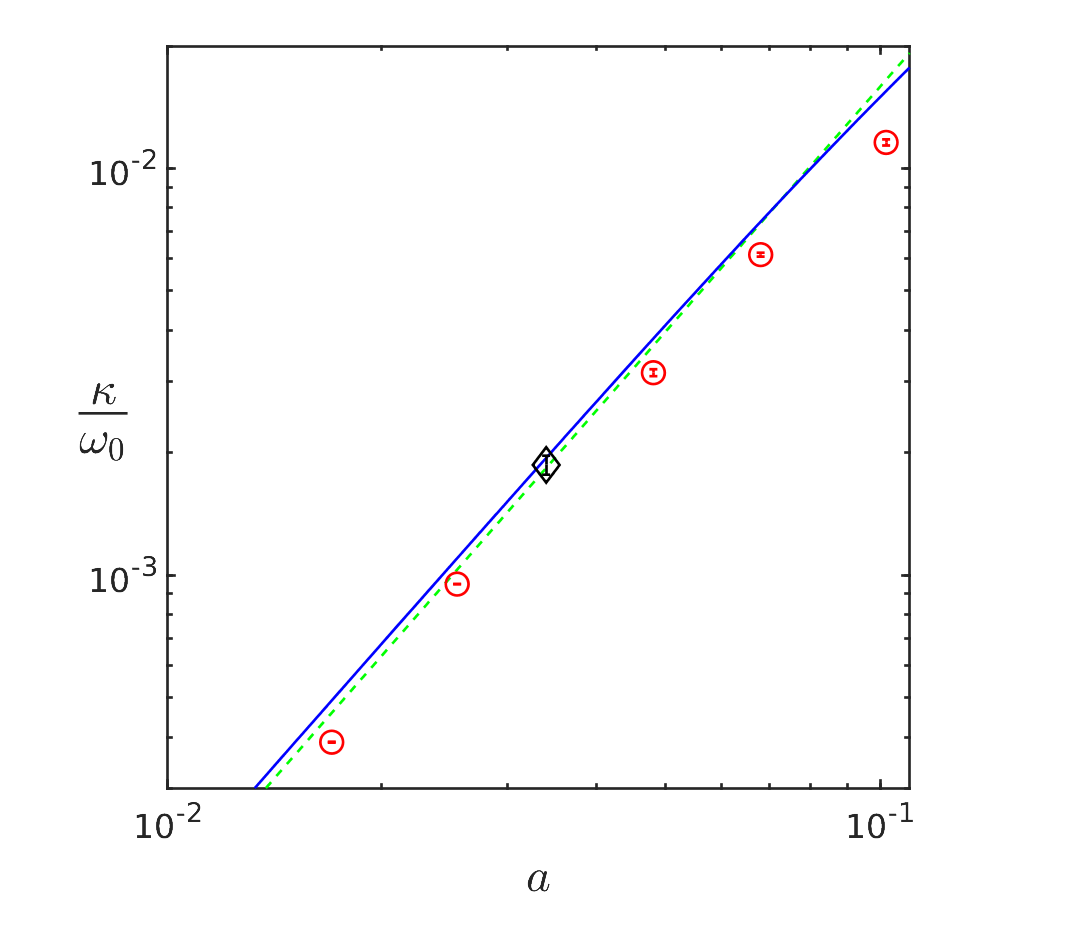}
    \caption{
    Induced scattering rates in nominal 1D simulations except a modified $T/m$ (left panel) or $a$ (right).
    Shown are the growth rates of $\delta S/S_0$ in our otherwise nominal 1D (induced scattering) simulations (red circles with fitting error bars),
    and of the fastest growing mode found by numerically solving
    Eq.~\eqref{eq:scatdisp} (blue solid) or from the asymptotic Eq.~(\ref{eq:finalgrowth}) (dashed green; valid under the condition Eq.~(\ref{eq:scatter_validity})).
    The black error bars inside diamonds
    demonstrate the combination of both 
    fitting and convergence uncertainties, seen to be quite small.
    \label{fig:modindscat}}
\end{figure*}

The simulated evolution of the induced scattering of a monochromatic strong wave is presented in Figs.~\ref{fig:densgrowth}--\ref{fig:modindscat}.
All figures are based on nominal 1D (induced scattering, nominal parameter, EPOCH) simulations, with two exceptions: the simulation of Fig.~\ref{fig:densgrowth} uses a very high $N_{ppc} = 1200$ (albeit a smaller $L_x = 100\lambda_0$) to resolve the density perturbations, and  Fig.~\ref{fig:modindscat} uses simulations where one physical parameter is varied, to present results for a wide range of physical parameters.

The temporal evolution of the different plasma components is shown in Fig.~\ref{fig:densgrowth}.
After an initial, oscillatory transient, the scattered electromagnetic modes are seen to grow exponentially, at a rate consistent with theory (nearly parallel measurement in solid red curve, theory in  thin dot-dashed magenta, and automatic fitting result in thin dashed black).
As anticipated above and shown numerically below, this growth is strongly dominated by the fastest growing mode, so a good agreement is obtained even when comparing the growth rate of all modes combined, $\delta S/S_0$, to the theoretical growth rate $\kappa_{\rm max}$ of the fastest growing mode.
The generation of scattered modes is accompanied by the growth of density perturbations (dotted teal).
To better resolve the growth in $\delta N_{\rm rms}/N_0$, we lower the respective noise level from $\sim 20\%$
to $\sim 0.7\%$ by removing small-scale noise using a high, $k>4k_0$ filter, as motivated by Fig.~\ref{fig:periodogram} below. The exponential growth of induced scattering modes is seen to saturate around $t\simeq 8000\omega_0^{-1}\simeq 15\kappa_{\rm max}^{-1}$. Subsequently, the electromagnetic modes still grow, but no longer exponentially, and the built-up density perturbations decline nearly exponentially.
The decay of the incident wave (dot-dashed cyan) and the heating of the plasma (dashed orange) accelerate after saturation, and are approximately exponential with a rate $\sim 3\times 10^{-5}\omega_0$.

The density perturbations can be seen at earlier times if one takes a power spectrum of the density field, as demonstrated in Fig.~\ref{fig:periodogram} for three characteristic stages: early linear phase ($\kappa_\mathrm{max}t = 7$), late linear phase ($\kappa_\mathrm{max}t = 15$) and saturation phase ($\kappa_\mathrm{max}t = 30$).
Even at early times, the figure shows pronounced beating at the expected angular wavenumber (magenta triangle), $k_d/k_0\simeq 2-2(T/m)^{1/2}\simeq 1.93$ (for nominal temperature); see Eq.~\eqref{eq:Backscattering}.
At later times, a second harmonic emerges, and a shift of both primary and second harmonic to slightly lower $k$ is seen to develop, consistent with plasma heating.
At small scales with $k\gtrsim 4k_0$ we find only numerical noise even at late times, justifying the high $k$ filter used in Fig.~\ref{fig:densgrowth}.
We verify in Fig.~\ref{fig:periodogram} (cyan thick curves) that fluctuations in electric charge $(N_p-N_e)/N_{\rm tot}$ remain small and are consistent with small-scale white noise.

To illustrate the scattered spectrum 
in more detail, Fig.~\ref{fig:specevo} presents the electromagnetic spectrum derived from a Fourier decomposition in traveling modes (see Appendix \ref{sec:expansion}).
The spectrum is shown for different times on the left panel, with an evolution plot (similar to Fig.~\ref{fig:densgrowth}) for reference in the right panel.
At early times, the back-scattered beam is seen to grow exponentially, sharply peaked at $k \simeq -0.9k_0$ wavelengths: consistent with the fastest growing mode being reversed and with a diminished, $\Delta\nu/\nu\simeq 2(T/m)^{1/2}$ frequency. After sufficient energy is transferred from the incident beam to the backscattered wave, the exponential growth saturates, and the reflected beam becomes broader in $k$-space.
The reflected beam itself is then back-scattered, leading to the growth of a broad-$k$ beam in the original direction, peaked around $k \simeq0.8k_0$.
At early times, there is a good agreement between the energy in the back-scattered mode and in all $k \neq k_0$ modes combined, but this is no longer true after saturation.
The figure demonstrates the basic picture of the induced scattering process (e.g., \citealt{Zeldovich_etal72}): radiation energy is distributed to lower frequencies while conserving the total number of photons, so the process necessarily involves some plasma heating.

To quantitatively compare the simulated mode growth with the induced scattering theory, Fig.~\ref{fig:dfcomp} shows the growth rate spectrum both as measured in the exponential phase of our nominal 1D simulation, and as derived asymptotically in Eq.~(\ref{eq:finalgrowth}).
For simplicity, we estimate the growth rate here as $\kappa(\bm{k})=\ln\left[S_{\bm{k}}\left(t_2\right)/S_{\bm{k}}\left(t_1\right)\right]/\left(t_2-t_1\right)$,
where the measurement time interval $[t_1,t_2]=[4,8]\kappa_\mathrm{max}^{-1}$ was chosen manually, and not by our nominal best-fit procedure.
Although no smoothing is used in this figure,
one can appreciate the good agreement between the simulated result (solid red curve) and the theoretical curve (dashed green).

Finally, we present in Fig.~\ref{fig:modindscat} a suite of simulations, showing how the induced scattering rate changes as a function of the plasma temperature (left panel) and the strength of the wave (right panel).
The measured rates from the otherwise nominal 1D simulations (error bars inside symbols) are compared to theory, showing both the numerical solution to the dispersion equation (\ref{eq:scatdisp}), and the asymptotic relation Eq.~\eqref{kappa_scat_max}.
Qualitatively, we find a good agreement between simulation and theory, within their respective limits, although we are comparing slightly different quantities: all $k\neq k_0$ modes outside the incident wave in the simulations, with the fastest growing mode in theory.
The red error bars inside circles show the fitting uncertainties of our rate measurements.
The black error bars inside diamonds show that even when combining fitting and convergence (from Richardson extrapolation) uncertainties, the resulting error estimate remains smaller than the approximation accuracy.

\subsubsection{Broad-spectrum beam}
\label{subsub:broadspec}

Next, consider the general case of a non-monochromatic incident beam, composed of many modes with random phases spanning a broad spectrum, as typically found in astrophysical applications.
We simulate the evolution of different such beams, defined as in \S\ref{subsec:InducedScattering} by a superposition of monochromatic waves, analogous to the distribution $a_\omega$ used \eg in Eq.~\eqref{eq:kappa_spectr}.
To simplify the comparison to the monochromatic case, we normalize the fields such that the total energy of any given beam equals that of the nominal monochromatic wave.

Consider a wave packet with a Gaussian energy distribution, centered on wavenumber $k_0$ with a standard deviation $\sigma$.
The implied injected fields are given by
\begin{equation}
B_z(x) = E_y(x) \propto \sum\limits_k
\cos \left( kx + \phi_k \right) e^{-\frac{\left(k-k_0\right)^2}{4\sigma^2}} \, ,
\label{eq:gaussian}
\end{equation}
where $\phi_k$ is a random phase,
and the summation is carried out over all $k$ modes up to the $|k-k_0|\simeq 7\sigma$ level around $k_0$. The factor $1/2$ introduced to the standard Gaussian exponent renders $\sigma$ the standard deviation in the energy, rather than in the field.

The extended spectrum of the beam should have a significant effect on the evolution of the plasma if it is not masked by the thermal spread, \ie when $\left(\sigma/k_0\right)^2 \gg T/m$. 
Numerically, one can represent the beam only by a finite, discrete set of modes, due to
the minimal,
$\delta k/k_0=\lambda_0/L_x$ spacing in $k$ space allowed by the periodic simulation box. To avoid numerical artefacts due to this non-continuum spectrum,
we require that the discreteness is masked by thermal broadening, $\left(\delta k/k_0 \right)^2 \ll T/m$. Combined, we thus require all simulated broad beams to satisfy
\begin{equation}
\label{widthineq}
    \left(\frac{\lambda_0}{L_x}\right)^2 =\left(\frac{\delta k}{k_0}\right)^2 \ll \frac{T}{m} \ll \left(\frac{\sigma}{k_0}\right)^2 \simeq \frac{\mathcal{N}^2}{4}\left(\frac{\delta k}{k_0}\right)^2 \, ,
\end{equation}
where $\mathcal{N} \simeq 2\sigma/\delta k$ is the number of modes injected within $1\sigma$ of $k_0$ in the numerical representation of the Gaussian in Eq.~\eqref{eq:gaussian}.
\begin{figure}[h!]
\begin{center}
\includegraphics[trim={0.5cm 0.45cm 1.5cm 0.1cm},clip,width=0.5\textwidth]{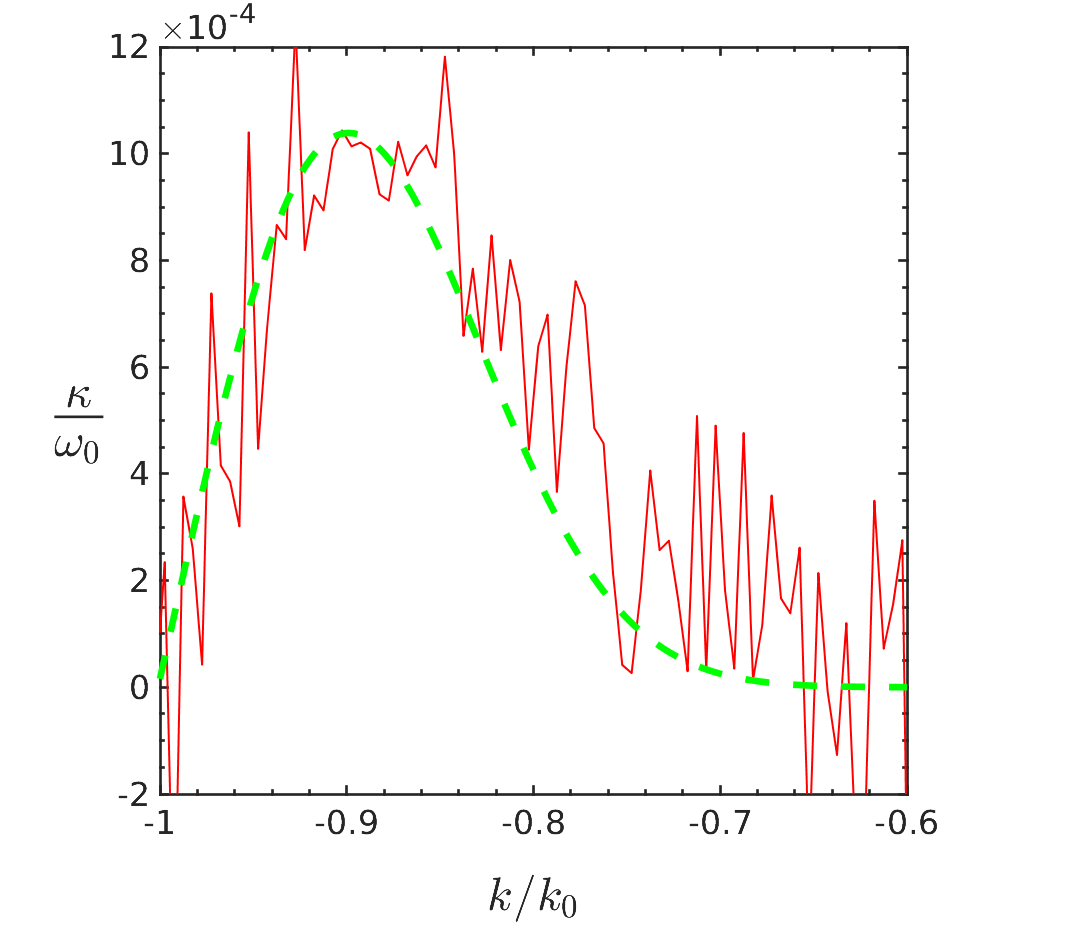}
\end{center}
\caption{The mean growth rate of different electromagnetic $k$-modes in a nominal 1D (induced scattering) simulation, with the monochromatic incident wave replaced by a broad beam with a Gaussian energy distribution of standard deviation $\sigma= 0.1k_0$ in $k$-space, during the exponential growth phase (time interval $[t_1,t_2]=[8,16]\kappa_\mathrm{max}^{-1}$; solid red), compared to the asymptotic theory (Eq.~\eqref{eq:kappa_spectr}; dashed green).
\label{fig:wpcomp}}
\end{figure}

We perform a suite of nominal 1D simulations, but replacing $a$ with a spectrum $a_\omega$ carrying the same total energy as in the monochromatic case. This spectrum is chosen as the Gaussian beam \eqref{eq:gaussian}, with a range of widths: $\sigma/k_0 \simeq 0.04$, $0.07$, $0.1$, $0.13$, $0.16$, $0.18$, and $0.21$.
Conditions \eqref{widthineq} are then satisfied, except possibly for the first few, narrow Gaussians, where
$\left(\sigma/k_0\right)^2/(T/m)\simeq 1.2,3.6,\ldots$, 
so the spectral extent of the beam may be partly washed out by the thermal broadening.
\begin{figure*}[t!]
     \centering
      \includegraphics[trim={1.0cm 0.5cm 2.0cm 0.5cm},clip,width=0.48\textwidth]{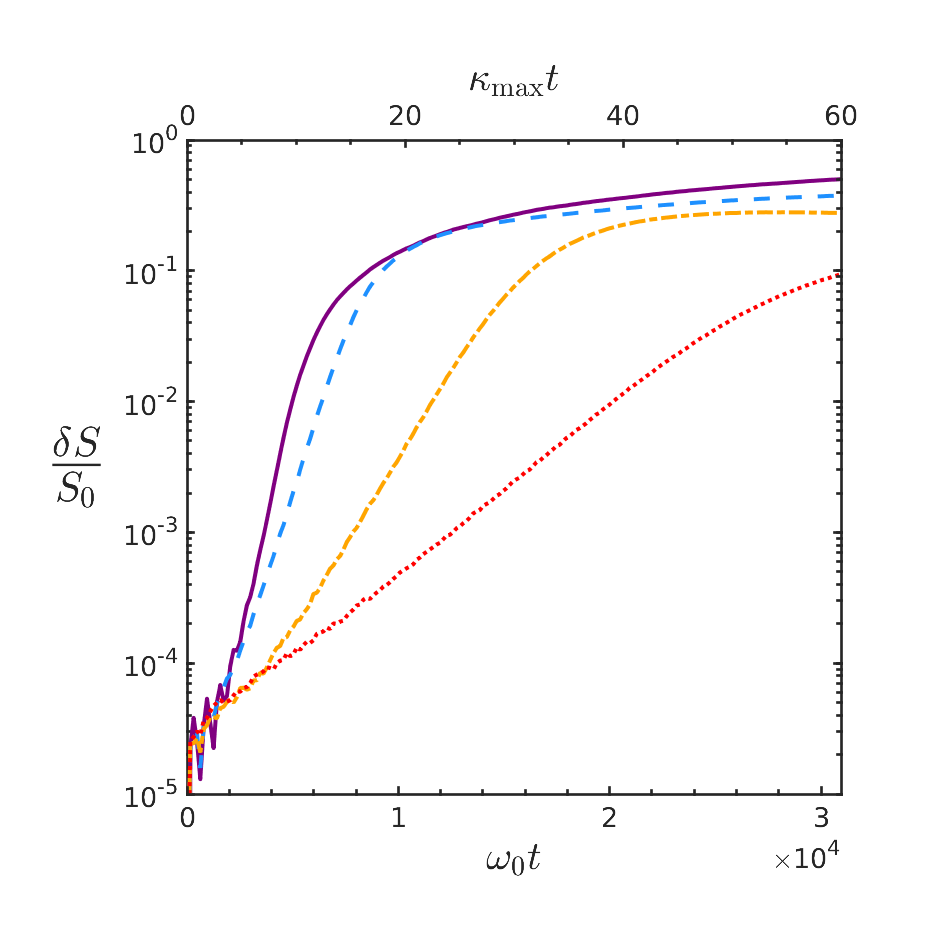}
      \includegraphics[trim={1.0cm 0.5cm 3.5cm 0.5cm},clip,width=0.48\textwidth]{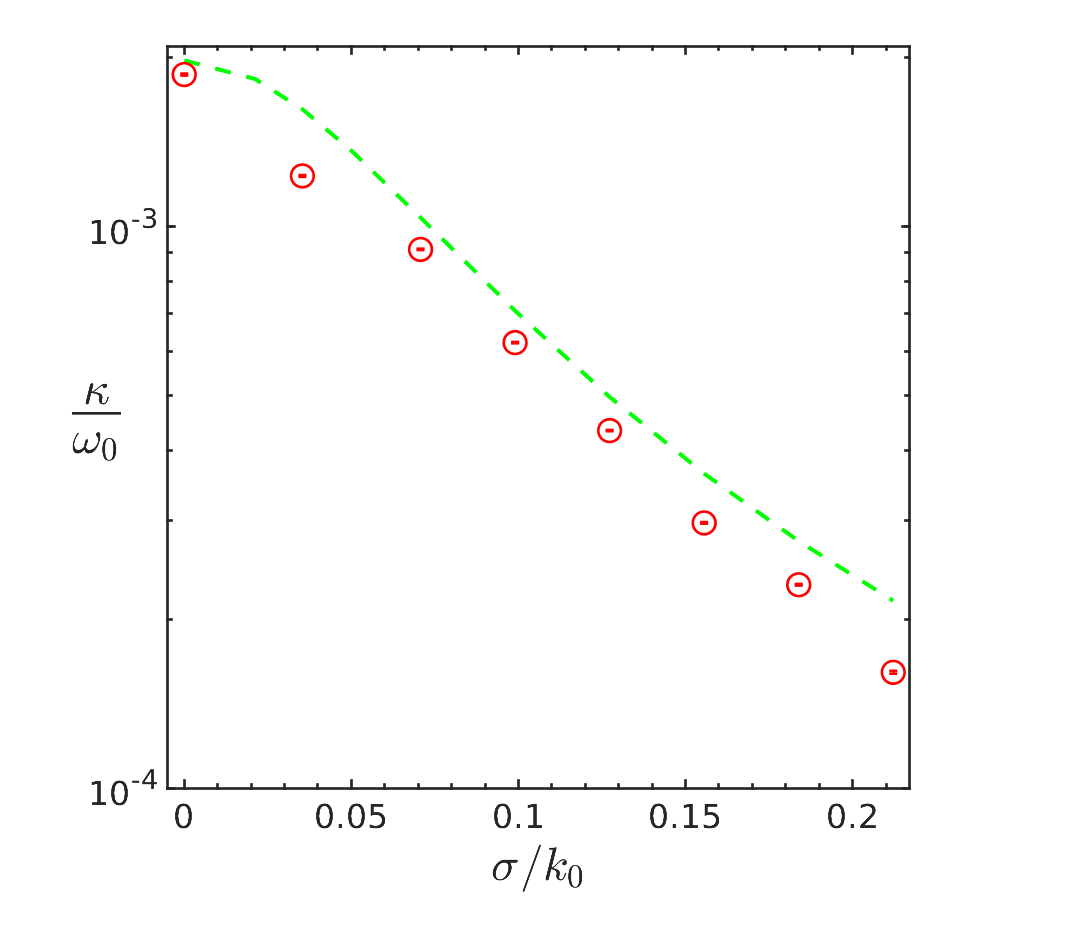}
    \caption{\label{fig:widegrowth}
    The induced scattering of broad beams with different Gaussian energy distributions, in otherwise nominal 1D simulations. \textit{Left:} Evolution of the electromagnetic flux outside the injected, $|k-k_0|<7\sigma$ beam, normalized to the initial injected flux, shown for four representative beams with standard deviations $\sigma/k_0 = 0$ (solid purple), $0.05$ (dashed blue), $0.1$ (dash-dotted yellow), and $0.16$ (dotted red).
     \textit{Right:} The estimated growth rates (of $\delta S/S_0$; red circles with fitting error bars) and asymptotic maximal rates (green dashed; obtained from Eq.~\eqref{eq:kappa_spectr}), as a function of $\sigma$.}
\end{figure*}

All these simulated broad beams yield growth rates that qualitatively agree with the theory and are modified with respect to the monochromatic case.
The case $\sigma=0.1k_0$ is demonstrated in Figure \ref{fig:wpcomp}, which shows the growth rate of different reflected $k$-modes.
Despite the noise obtained in the absence of any smoothing, the results are seen to be in good agreement with the asymptotic growth rate profile.
This profile is computed numerically from Eq.~\eqref{eq:kappa_spectr}, which convolves the monochromatic rate Eq.~\eqref{eq:finalgrowth} in the slow-growth limit \eqref{eq:conditionscat} with the spectrum of the beam.

To analyze the temporal evolution of the broad-beam simulations, we use the generalized definition of $\delta S/S_0$ as the electromagnetic flux outside the injected, $|k-k_0|<7\sigma$ beam, normalized to the initial injected flux.
The time-dependence of this quantity is shown in Figure \ref{fig:widegrowth} (left panel) for four representative values of $\sigma/k_0$.
Extended exponential growth phases are resolved, so the growth rates can be accurately measured. Note the irregular changes in saturation phase behavior as a function of $\sigma$.

The measured exponential growth rates of this generalized $\delta S/S_0$ are shown in the right panel of
Figure \ref{fig:widegrowth}, for all  simulations, including both broad beams and the monochromatic, $\sigma/k_0 = 0$ case.
These simulated rates are seen to be in qualitative agreement with the maximal theoretical growth rates inferred from the asymptotic Eq.~\eqref{eq:kappa_spectr}.
As expected, the rate declines with increasing $\sigma$.

\subsection{Filamentation instability}
\label{subsec:FilamentationResults}

We now turn our attention to the filamentation instability. A strong electromagnetic wave is injected into the pair plasma as in the
induced scattering case, but here it is necessary to resolve long-wavelength perturbations in the perpendicular, $y$-direction.
Hence, we use a 2D box with a large, $L_y=100\lambda_0$ transverse extent.
Computation resource limitations then dictate using lower values of $N_{ppc}$ and $\lambda_0/\Delta x$ (see \S\ref{subsec:Convergence}), and the minimal $L_x=\lambda_0$ possible still able to sustain the incident wave.

\begin{figure}[h!]
\centering
\includegraphics[width=0.48\textwidth]{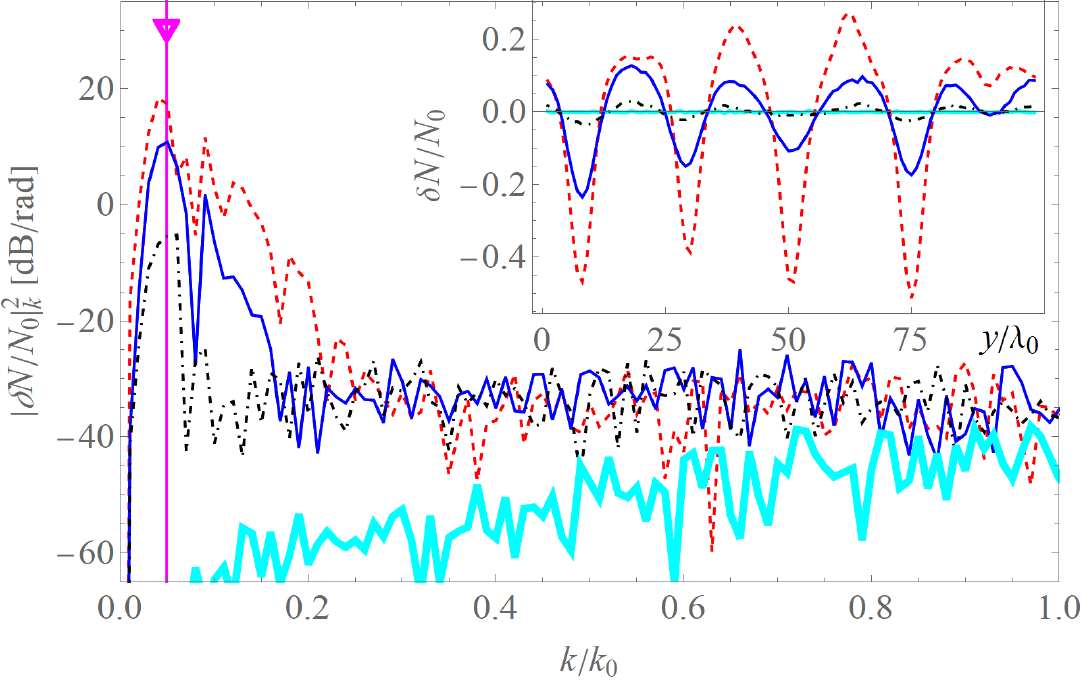}
\caption{Density perturbations (thin curves) from a nominal 2D (filamentation) simulation, shown during early
($t \simeq 5.7\times 10^3 ~\omega_0^{-1}\simeq 4\Gammamax^{-1}$; 
dot-dashed black) and
late ($t \simeq 8.6\times10^3 ~\omega_0^{-1} \simeq 6\Gammamax^{-1}$; solid blue) exponential growth, and during saturation ($t \simeq 1.1\times 10^4 ~\omega_0^{-1} \simeq 8\Gammamax^{-1}$; dashed red).
Filaments are seen forming in the density field (inset) and as the pronounced peak in the power spectrum (unfolded; figure body), consistent with the fastest growing mode given by Eq.~\eqref{eq:FilamentationFastestMode} (magenta triangle and vertical line). Filament merger and spectral broadening are seen to develop at late times.
Fluctuations in electric charge $(N_p-N_e)/N_{\rm tot}$ (thick cyan, for the later time) remain small and consistent with small-scale white noise.
\label{fig:Filaments}
}
\end{figure}

Normalized perturbations in the density profile $N(y)$, after averaging along $x$, are shown in Fig.~\ref{fig:Filaments} (inset) for the nominal 2D simulation.
An oscillatory mode of wavelength $\sim 100\lambda_0/6\simeq 17\lambda_0$ is seen to grow exponentially, evolving into dense, merging filaments separated by low-density plasma as saturation is approached.
Indeed, a power-spectrum of $N(y)$ (figure body) shows a fairly sharp peak at $k\simeq 0.06k_0$ early on, progressively broadening and shifting to lower $k$ at late times.
This wavenumber is broadly consistent with the $k_d\simeq (\omega_p a_0/2)(T/m)^{-1/2}\simeq 0.05k_0$ estimate of the fastest growing mode in Eq.~\eqref{eq:FilamentationFastestMode}, for our nominal physical parameters.
Note that Eq.~\eqref{eq:FilamentationFastestMode} should not be accurate here as it depends on the assumption $\omega_d\ll k_d v_T$, or equivalently Eq.~\eqref{eq:conditionFilament}, which is only weakly satisfied for these parameters; 
solving Eq.~\eqref{eq:disp1} numerically gives an even lower, $k \simeq 0.04k_0$.
The figure also shows that even at late times, there is only noise at small, $k\gtrsim k_0/3$ scales, and that the plasma remains neutral at all times.

\begin{figure}[h!]
\begin{center}
\includegraphics[trim={0.5cm 1.2cm 1.5cm 0.8cm},clip,width=0.5\textwidth]{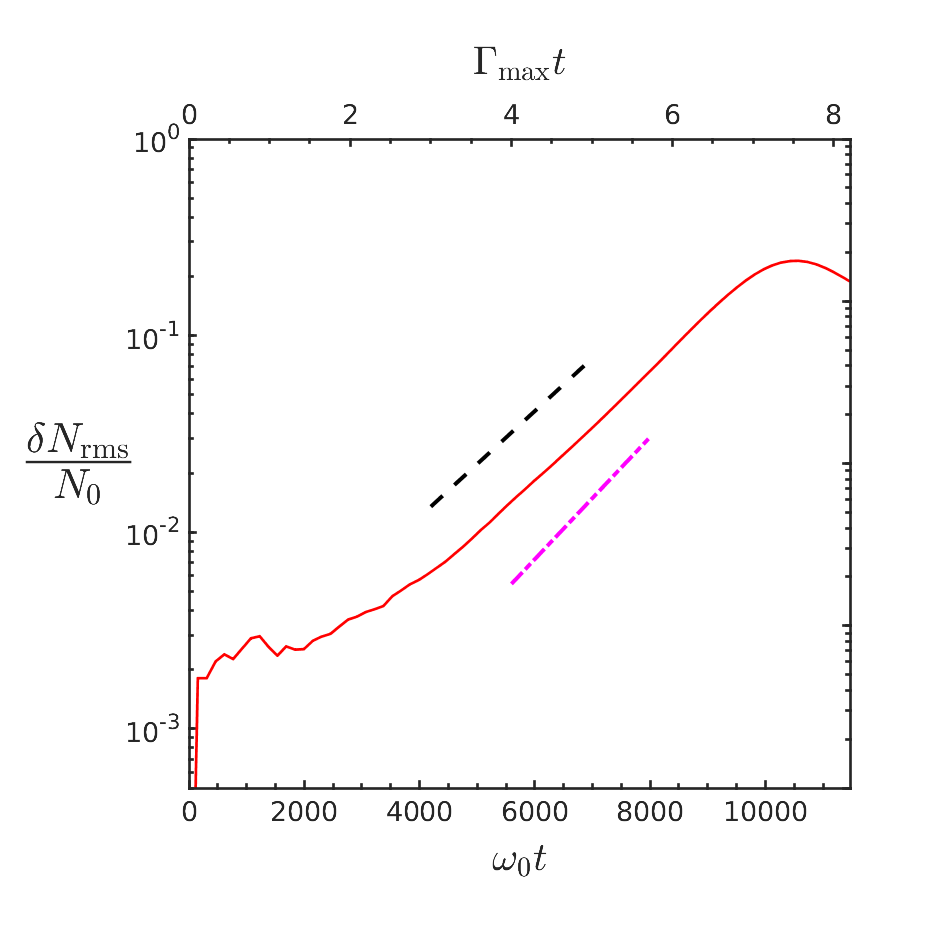}
\end{center}
\caption{Evolution of density perturbations along $y$ in a nominal 2D (filamentation) simulation.
The measured $\left(\delta N_\mathrm{rms}/N_0\right)$ is shown (solid red curve) after averaging over $x$ and removing short-wavelength, $k>k_0/3$ modes to lower the noise level.  The perturbations undergo a period of exponential growth, in agreement with the theoretical fastest growth rate (magenta dot-dashed slope; obtained by solving Eq.~\eqref{eq:disp1} numerically); our automatic fitting measurement is also shown (black dashed slope).
\label{fig:filagrowth}
}
\end{figure}

\begin{figure*}[t!]
     \centering
     \includegraphics[trim={0.0cm 0.5cm 2.5cm 0.5cm},clip,width=0.48\textwidth]{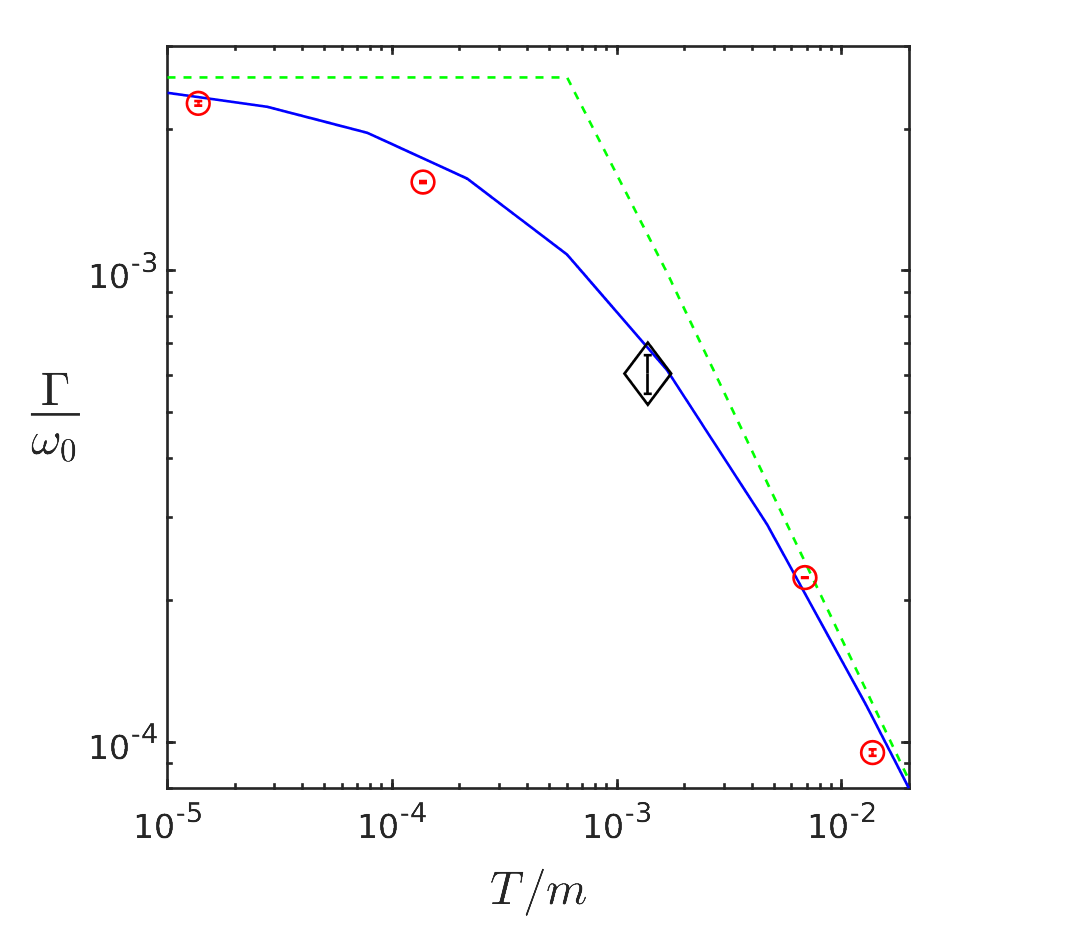}
     \includegraphics[trim={0.0cm 0.5cm 2.5cm 0.5cm},clip,width=0.48\textwidth]{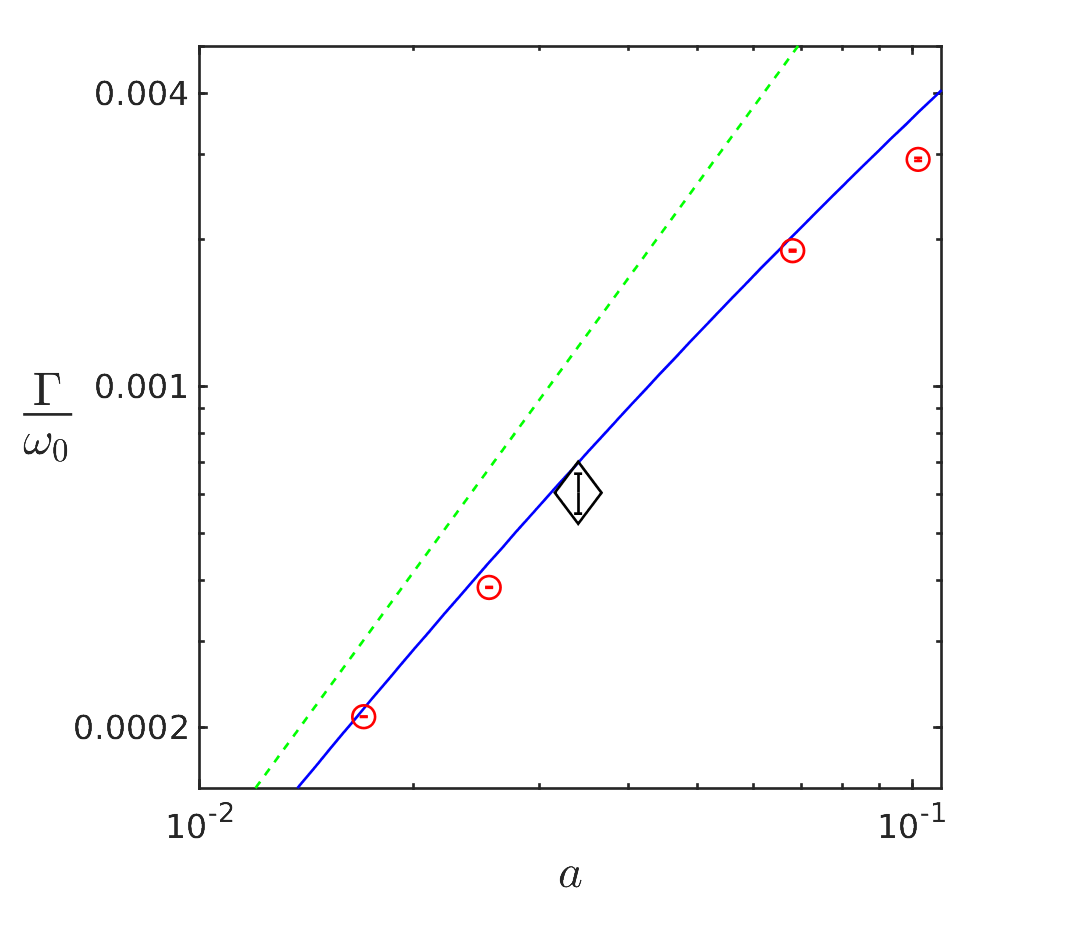}
        \caption{
        \label{fig:modfil}
        Filamentation growth rates obtained by varying $T/m$ (left panel) or $a$ (right) in otherwise nominal 2D (filamentation) simulations.
        The measured rates (red fitting error bars inside circles) are based on the RMS density perturbations, $\delta N_\mathrm{rms}/N_0$, after averaging along $x$ and filtering out all $k>k_0/3$ modes.
        The results are compared to the theoretical maximal growth rates, calculated by solving Eq.~\eqref{eq:disp1} numerically (solid blue curve) or using the asymptotic estimates (dashed green; minimum of the two rates in Eqs.\ (\ref{eq:FilamentationGrowthRate}) and (\ref{eq:filamentMaxGrowth1})). The black error bars inside diamonds, demonstrating the combination of both fitting and convergence uncertainties, are seen to be quite small.
        }
\end{figure*}

The temporal evolution of the RMS density fluctuations in a nominal 2D simulation is shown in Fig.~\ref{fig:filagrowth}. In order to lower the noise level and resolve a more prolonged period of exponential growth, here we first apply a high-$k$ filter, removing all $k>k_0/3$ modes which are indicated as noise by Fig.~\ref{fig:Filaments}.
The exponential growth rate is seen to be in good agreement with the fastest growth rate found by solving Eq.~\eqref{eq:disp1} numerically (nearly parallel measurement in solid red curve, theory in dot-dashed magenta, and automatic fitting result in dashed black).

Finally, we carry out a suite of simulations to investigate the filamentation instability as a function of physical parameters.
Figure \ref{fig:modfil} shows the dependence of the estimated growth rate $\Gamma$ of the instability upon $T/m$ (left panel) and $a$ (right panel).
These rates are compared to the fastest growth rate derived from the numerical solution to the dispersion relation Eq.~\eqref{eq:disp1}, as well as the approximate asymptotic relations given by Eqs.~\eqref{eq:displow} and \eqref{eq:disphigh}. One sees that the growth rates obtained from the simulations agree very well with the theory, over a wide range of physical parameters.

\section{Summary}
\label{sec:summary}

This is the first paper in a series devoted to the study of nonlinear interactions of powerful electromagnetic radiation with pair plasma. These effects could play a role in different astrophysical environments, including AGN, pulsars and interstellar masers, and drew considerable interest in recent years, after the discovery of fast radio bursts. In this paper, we study the case where the strength parameter of the wave, defined in Eq.\ (\ref{eq:a}), is small, $a\ll 1$. Here, particles oscillate in the field of the wave with velocities well below the speed of light. This case can thus be studied analytically, facilitating the development of numerical tools by comparing the theoretical and numerical results.

At the initial stage, perturbations caused by the strong radiation beam grow exponentially, and the evolution may be considered as an instability of the beam. Therefore, the first question is what are the characteristics and growth rates of the relevant instabilities. The two main effects for a pair plasma are induced scattering and the filamentation instability. We derived dispersion relations for these instabilities in Eqs.~\eqref{eq:scatdisp} and \eqref{eq:disp1}, respectively, from which the corresponding growth rates are easily calculated numerically. The well-known asymptotic relations in different limiting cases are easily reproduced from these dispersion relations.

We performed extensive numerical simulations to study the temporal evolution of an initially homogeneous pair plasma exposed to powerful radiation, and measured the properties of the resulting instabilities. The simulations were carried out using both EPOCH and Tristan-MP, with good agreement between the two codes, and with small statistical and convergence uncertainties (see Appendix \ref{app:Convergence} and Figs.~\ref{fig:compgrowth}--\ref{fig:grconv}).
By choosing long, narrow simulation boxes either parallel ($L_x\gg L_y$, and nominally with $L_y=0$) or perpendicular ($L_y\gg L_x$) to the incident beam, we select for either induced scattering or filamentation growth, allowing us to study each effect separately. The results, assisted by an algorithm for identifying the exponential growth period (\S\ref{subsec:GrowthRates}) and a decomposition of fields in traveling waves (Appendix \ref{sec:expansion}), are shown in Figs.~\ref{fig:densgrowth}--\ref{fig:Filaments} for induced scattering, and in Figs.~\ref{fig:filagrowth}--\ref{fig:modfil} for the filamentation instability.

The simulations show the initial exponential growth of perturbations in electromagnetic fields and in density, and the transfer of energy from the incident  beam to lower frequencies and to particle oscillations, in good quantitative agreement with theory for a wide range of physical parameters (see in particular Figs.~\ref{fig:modindscat} and \ref{fig:modfil}) and across multiple diagnostics. The simulations trace the evolution deep into the saturation phase, showing non-linear structure growth, the dissipation of the density perturbations, and substantial plasma heating. For the case of induced scattering, where the spectrum of the beam is important, we examine both monochromatic and broad beams.
The growth becomes slower as the incident spectrum is broadened, in agreement with theory, implying that for astrophysically relevant beams, filamentation occurs well before the induced scattering comes into the play. 

Having established consistent theory and simulations in the $a \ll 1$ regime, with different codes and multiple diagnostics, future papers in this series will explore the strong $a$ regime, the addition of background magnetic fields, the inclusion of ions, etc.

\section*{Acknowledgments}

This research was supported by grant I-1362-303.7/2016 from the German-Israeli Foundation for
Scientific Research and Development, by the Israeli Science Foundation (grants 2067/19 and 1769/15), and by the IAEC-UPBC joint research foundation (grant No. 300/18).

\appendix
\twocolumngrid

\section{Nonlinear current: monochromatic wave}
\label{sec:current}

Here we briefly outline the derivation of the nonlinear current in a plane monochromatic wave \citep{Montgomery_Tidman64,Sluijter_Montgomery65}. The current is produced by electrons oscillating in the field of the wave, see Eq.\ (\ref{eq:current}). In the first approximation, the oscillation velocity is given  by Eq.\ (\ref{eq:velocity}).
Specifically, for a plane monochromatic wave, \begin{equation}
\mathrm{e}\bm{A}/m=a\hat{\bm{y}}\cos\eta;\quad \eta=\omega_0 t-k_0x,
\end{equation}
one gets
\begin{equation}
\bm{v}^{(1)}=-\frac{q}{\mathrm{e}} a\hat{\bm{y}}\cos\eta. \label{eq:v_y}\end{equation}
To first approximation, this yields the linear current (\ref{linear_currrent}).

The next approximation is of the third order in $a$. Two effects contribute to the current in this order: the density variation in the field of the wave,
\begin{equation}
\bm{j}_1^{\rm nl}=\sum q\,\delta N^{(2)}\bm{v}^{(1)},
\label{eq:j1}\end{equation}
and the nonlinear corrections to the particle oscillation velocity,
\begin{equation}
    \bm{j}_2^{\rm nl}=N_0\sum q \bm{v}^{(3)}.
\label{eq:j2}\end{equation}
Therefore, one can conveniently present the nonlinear current as a superposition of two terms.

The first contribution arises from the Lorentz force (the second term in the RHS of Eq.\ (\ref{eq:motion})).
Due to oscillations with velocity (\ref{eq:v_y}), the magnetic field of the wave produces a longitudinal (along the direction of the pumping wave) force at the doubled frequency,
 \begin{equation}
     q\bm{v\times (\nabla\times A)}=-\frac 12mk_0a^2\hat{\bm{x}}\sin2\eta.
     \label{eq:LongitudinalForce}
 \end{equation}
This force yields longitudinal oscillations in the second order in $a$:
\begin{equation}
    \bm{v}^{(2)}=\frac 14\frac{k_0a^2}{\omega_0}\hat{\bm{x}}\cos2\eta.
\end{equation}
These oscillations produce, by virtue of the continuity equation, the density perturbation, $\delta N^{(2)}=(k_0/\omega_0)N_0v_{x}^{(2)}$. The beating between the density oscillations at the double frequency and the velocity oscillations (\ref{eq:v_y}) contributes to the nonlinear current (\ref{eq:j1}). Taking into account that $\delta N$ is a quadratic function of the particle charge, so that the electron and positron fluctuations are equal, one gets
\begin{gather}
    4\pi\bm{j}^{\rm nl}_{1}=-\frac{ma^3k_0^2}{8\mathrm{e}\omega_0^2}\omega_p^2\left(\cos3\eta+\cos\eta\right)\bm{\hat y}.
\end{gather}
The second contribution to the nonlinear current arises because relativistic corrections effectively raise the mass of an oscillating particle. In the plane wave, the component of the generalized momentum perpendicular to the propagation direction is conserved, $p_y+qA_y=\const$; one can check directly that this relation satisfies the equation of motion (\ref{eq:motion}) exactly. In our case, this relation reduces to
\begin{equation}
    \frac{v_y}{\sqrt{1-v^2}}+\frac{q}{\mathrm{e}} a\cos\eta=0.
\end{equation}

In the first approximation, this relation yields Eq.\ (\ref{eq:v_y}). In the next approximation, 
one finds that $v_{y}$ has a correction third order in $a$:
\begin{equation}
    v_y^{(3)}+\frac 12 \left(v_y^{(1)}\right)^3=0 \,.
\end{equation}
This yields the second contribution to the non-linear current, Eq.\  (\ref{eq:j2}): 
\begin{equation}
     4\pi\bm{j}^{\rm nl}_2
     =\frac{ma^3}{8\mathrm{e}}\omega_p^2\left(\cos3\eta+3\cos\eta\right)\bm{\hat y}.
\end{equation}
Together, $j^{\rm nl}_1$ and $j^{\rm nl}_2$ give the nonlinear current. However, only resonant terms (those oscillating with the frequency $\omega_0$) should be taken into account in the wave equation because only perturbations from these terms accumulate at the large scale whereas nonresonant terms produce small perturbations that do not grow. Retaining only resonant terms yields
\begin{equation}
  4\pi  \bm{j}^{\rm nl}_{\rm res}=
    \frac 18a^2\omega_p^2\left(3-\frac{k_0^2}{\omega_0^2}\right)\bm{A},
    \end{equation}
which gives
Eq.\ (\ref{eq:nl_current}) in the high frequency limit, $k_0=\omega_0$.

\section{Expanding the fields in travelling waves}
\label{sec:expansion}

Simulation snapshots provide the discretized fields $\bm{E}(\bm{r})$ and $\bm{B}(\bm{r})$ at each time step.
One could expand the fields in Fourier series,
\begin{equation}
\begin{aligned}
\bm{E}(\bm{r})=\sum_{\bm{k}} \bm{e}_{\bm k}e^{i \bm{k \cdot  r}}; \\
\bm{B}(\bm{r})=\sum_{\bm{k}} \bm{b}_{\bm k}e^{i \bm{k \cdot  r}},
\end{aligned}
\label{eq:Fourier}
\end{equation}
where $\bm{e}_{\bm k}=\bm{e}_{-\bm k}^*$ and $\bm{b}_{\bm k}=\bm{b}_{-\bm k}^*$ because $\bm{E}$ and $\bm{B}$ are real.
However, such an expansion does not facilitate the separation of waves propagating in opposite directions. In order to analyze the results of simulations of the scattering process, we expand the fields in travelling waves,
\begin{equation}
\begin{aligned}
\bm{E}(\bm{r},t)=\sum_{\bm{k}} \bm{E}_{\bm k}e^{i( {\bm{k \cdot r}-\omega t})}+ {\rm c.c.};\\
\bm{B}(\bm{r},t)=\sum_{\bm{k}} \bm{B}_{\bm k}e^{i( \bm{k \cdot r}-\omega t)}+ {\rm c.c.}.
\end{aligned}
\label{eq:travelling_waves}
\end{equation}

In this case, the direction of $\bm{k}$ gives the direction of the wave. The Poynting flux is presented as
\begin{equation}
    \bm{S}=\frac{1}{4\pi}\langle\bm{E\times B}\rangle=
    \frac{1}{4\pi} \sum_{\bm{k}} \bm{E_{k}\times B_{k}}^*+{\rm c.c.},
\end{equation}
so that the flux of each wave is
\begin{equation}
    \bm{S_k}=
    \frac{\bm{E_{k}\times B_{k}}^*+
    \bm{E_{k}}^*\bm{\times B_{k}}}{4\pi}.
\label{eq:Poynting}\end{equation}

By comparing Eqs.\ (\ref{eq:Fourier}) and (\ref{eq:travelling_waves}), we find
\begin{equation}
\begin{aligned}
\bm{e}_{\bm k}= {\bm E}_{\bm k}+{\bm E}^{*}_{-\bm k};\\
\bm{b}_{\bm k}= {\bm B}_{\bm k}+{\bm B}^{*}_{-\bm k}.
\end{aligned}
\end{equation}
Here we take, without loss of generality, $t=0$.
These two equations are supplemented by two Maxwell equations; for the travelling waves (\ref{eq:travelling_waves}) they are written as
\begin{equation}
\omega {\bm B}_{\bm k}= {\bm k}\times  {\bm E}_{\bm k};\quad
 {\bm k}\cdot  {\bm E}_{\bm k} =0.
\end{equation}
Now we have four equations, so we may express the amplitudes of the travelling waves via the Fourier components of the snapshots:
\begin{align}
{\bm E}_{\bm k} &=\frac{{\bm e}_{\bm k}- \hat{\bm k} \times {\bm b}_{\bm k}}{2}\,;\\
{\bm B}_{\bm k} &=\frac{{\bm b}_{\bm k}+\hat{\bm k} \times {\bm e}_{\bm k}}{2}\,,
\label{eq:btime}
\end{align}
where $\hat{\bm k}$ is the unit vector in the direction of the wave, and we take into account that $\omega=k$ for high-frequency waves. Substituting these relations in Eq.\ (\ref{eq:Poynting}), we get the Poynting flux in each wave as
\begin{equation}
\bm{S_ k} =\frac{\vert{\bm e_{\bm k}}\vert^2+ \vert{\bm b_{\bm k}}\vert^2}{8\pi}\hat{\bm k}+
\frac{\bm{e}_{\bm{k}}\times\bm{b}_{\bm{k}}^*+\mbox{c.c.}}{8\pi}\, .
\label{eq:SkFourier} \end{equation}

\section{Simulation Convergence}
\label{app:Convergence}

The numerical parameters associated with our PIC simulations are $N_\mathrm{ppc}$, 
$\lambda_0/\Delta x$, 
and either $L_x/\lambda_0$ (for 1D simulations) or $L_y/\lambda_0$ (for 2D). 
Convergence tests based on nominal 1D simulations 
demonstrate the convergence of the growth rates inferred from our automatic fitting procedure (see \S\ref{subsec:GrowthRates}), with respect to all numerical parameters, using a large suite of simulations.
For example, one test examines the convergence of the results with a control parameter $\mathcal{C}$, which scales both $N_\mathrm{ppc}$ and $\lambda_0/\Delta x$ simultaneously, thus multiplying the number of particles per wavelength by $\mathcal{C}^2$.

\begin{figure}[h!]
\begin{center}
\includegraphics[trim={0.5cm 0.5cm 2.5cm 0cm},clip,width=0.48\textwidth]{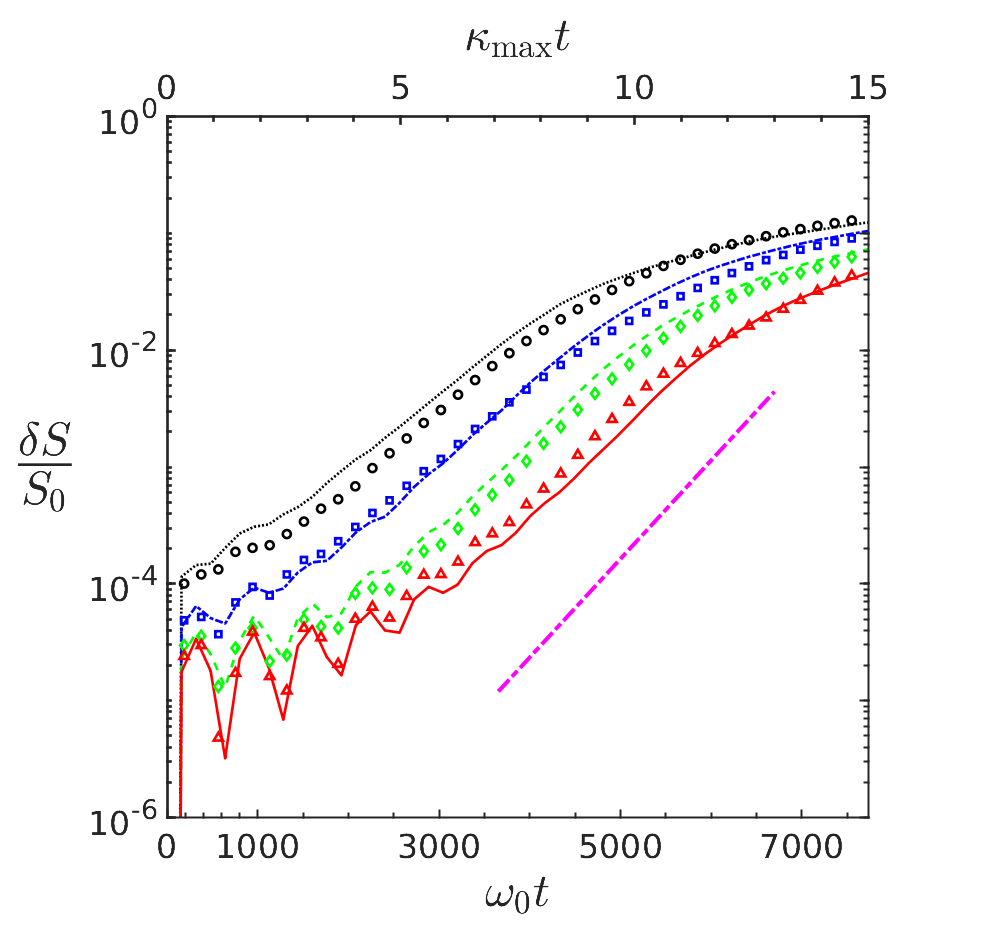}
\caption{
\label{fig:compgrowth}
Convergence and agreement of induced scattering simulations using EPOCH (lines) and Tristan-MP (markers, with a smaller, $L_x = 100\lambda_0$).
The electromagnetic energy fraction $\delta S/S_0$ of $k\neq k_0$ modes 
is shown for nominal parameters after multiplying both $N_\mathrm{ppc}$ and $\lambda_0/\Delta x$ by the same factor
$\mathcal{C}=0.25$ (black circles and dotted curve),
$0.5$ (blue squares and dot-dashed),
$1$ (nominal; green diamonds and dashed),
and
$2$ (red triangles and solid).
The theoretical fastest growth rate is also shown (thin magenta dot-dashed slope). 
}
\end{center}
\end{figure}

\begin{figure*}[t!]
    \centering
    \includegraphics[trim={0.5cm 0.5cm 2.5cm 0.0cm},clip,width=0.48\textwidth]{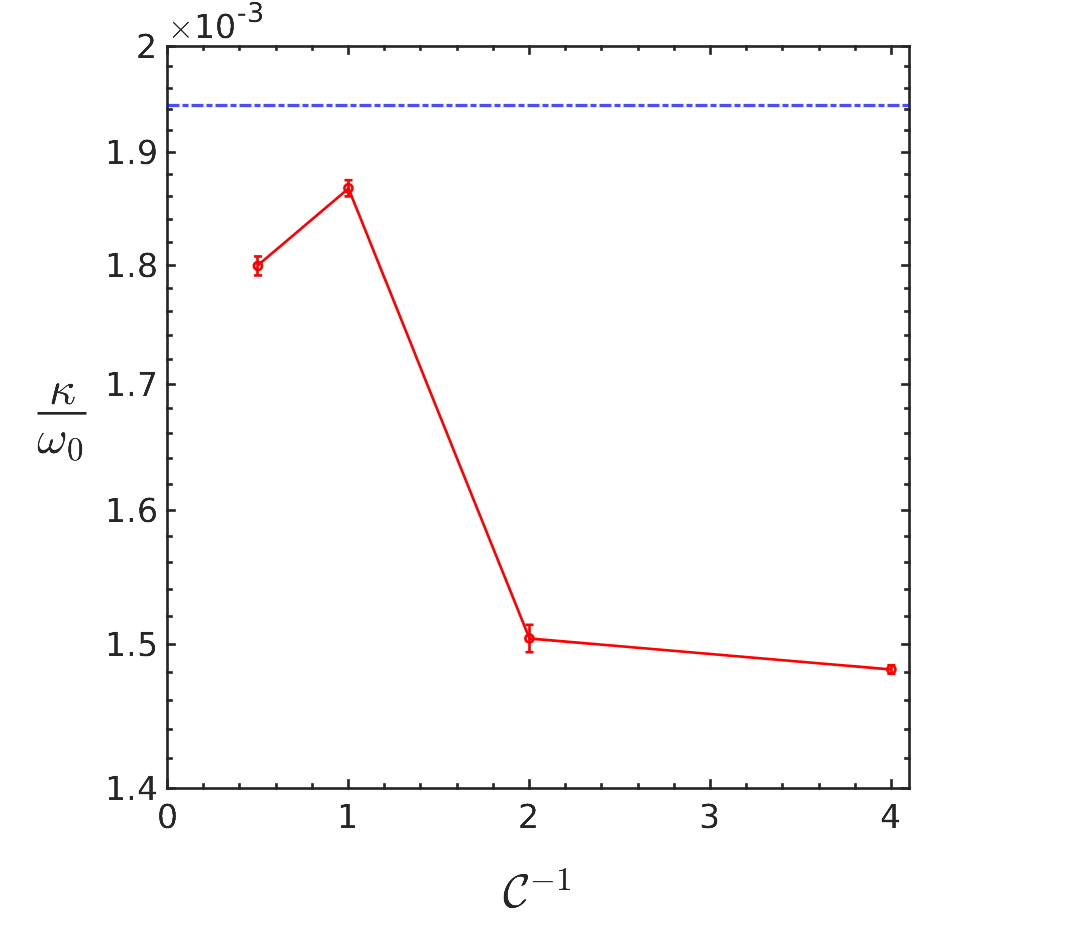}
    \includegraphics[trim={0.5cm 0.5cm 2.5cm 0.0cm},clip,width=0.48\textwidth]{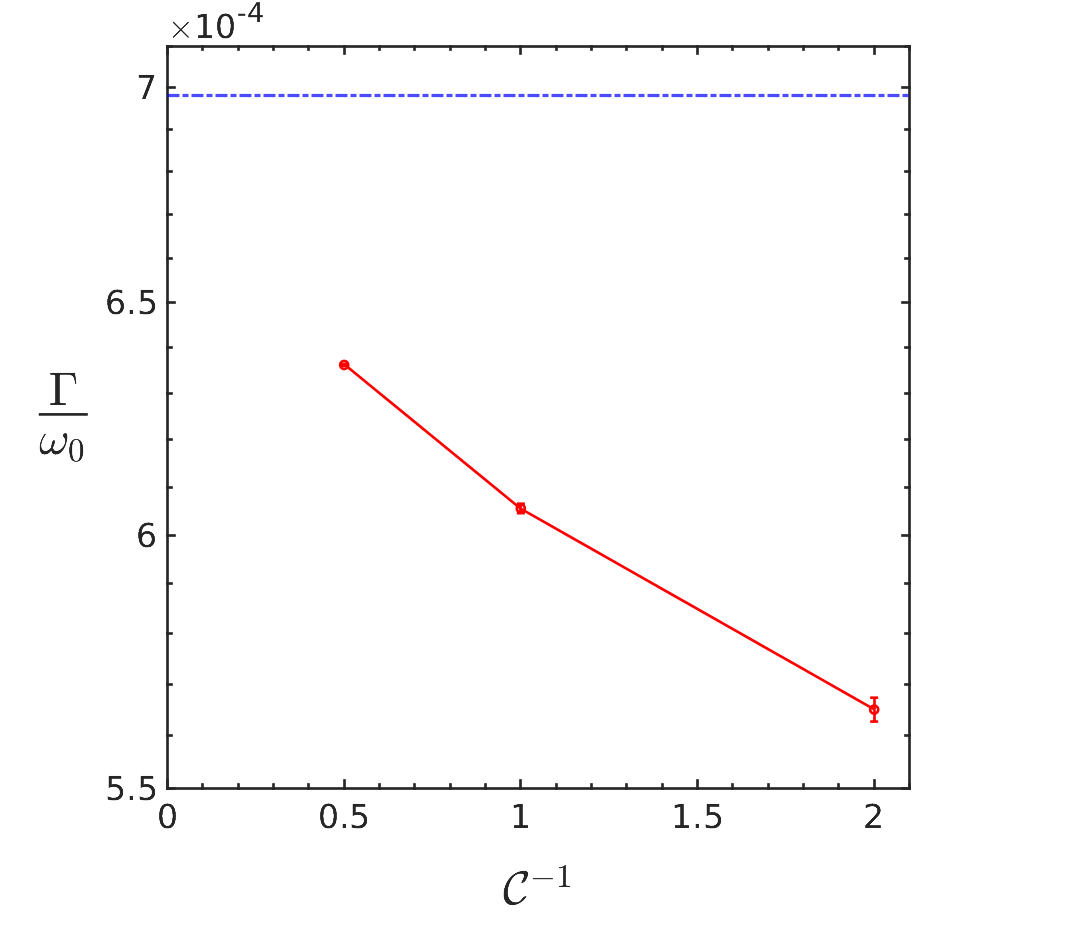}
    \caption{Convergence test based on nominal 1D (left) and 2D (right) simulations, after multiplying $N_\mathrm{ppc}$ and $\lambda_0/\Delta x$ by the same factor $\mathcal{C}$. 
    Shown are the automatically fitted growth rates (see \S\ref{subsec:GrowthRates}; red error bars connected by lines to guide the eye) and the theoretical growth rate (blue dash-dotted lines, calculated from Eqs.~\eqref{eq:scatdisp} and \eqref{eq:disp1}). 
    \label{fig:grconv}}
\end{figure*}

We test the agreement between EPOCH and Tristan-MP results using a suite of 1D simulations, based on the nominal parameters but modified by different values of $\mathcal{C}$, where $\mathcal{C}=1$ is the nominal case. Figure \ref{fig:compgrowth} shows the evolution 
of the fractional electromagnetic perturbations 
for four different choices of $\mathcal{C}$, in both EPOCH (lines) and Tristan-MP (markers).
The growth rates inferred from the different simulations are consistent with each other and with the theoretical value (magenta dot-dashed slope). Figure \ref{fig:grconv} demonstrates the convergence of the fitted growth rates obtained as a function of the scaling factor $\mathcal{C}$ in our otherwise nominal 1D (left panel) and 2D (right panel) simulations.
Richardson extrapolation to $\mathcal{C}\to0$ suggests convergence uncertainties of order $\lesssim 15\%$. 

\bibliographystyle{apj}
\bibliography{FRB}

\label{lastpage}	
\end{document}